\def\mum {\hbox{$\mu$m}}

\def\cii{\hbox{{\rm [C {\scriptsize II}]}}}
\def\ci{\hbox{{\rm [C {\scriptsize I}]}}}
\def\oi{\hbox{{\rm [O {\scriptsize I}]}}}

\newcommand{\hii}{H{\sc ii}}

\documentclass{aa}  
\usepackage{natbib}
\usepackage{float}
\usepackage{placeins}
\usepackage{graphicx}
\usepackage{gensymb}
\usepackage{multirow}
\usepackage{txfonts}
\usepackage{subfigure}
\usepackage{fixltx2e}
\usepackage[colorinlistoftodos]{todonotes}
\usepackage[flushleft]{threeparttable}
\usepackage[normalem]{ulem}
\usepackage{dcolumn}
\usepackage{csquotes}

\newcolumntype{d}[1]{D{.}{\cdot}{#1}}
\newcolumntype{.}{D{.}{.}{-1}}

\begin{document}

   \title{Observational study of hydrocarbons in the bright photodissociation region of Messier 8}

   \subtitle{}

   \author{M.\,Tiwari\inst{1}\fnmsep\thanks{Member of the International Max Planck Research School (IMPRS) for Astronomy and Astrophysics at the Universities of Bonn and Cologne.}\fnmsep, K.\,M.\,Menten\inst{1}, F.\,Wyrowski\inst{1}, J.P.\,P\'{e}rez-Beaupuits\inst{2,1}, M.\,-Y.\,Lee\inst{1}, W.-J.\,Kim\inst{3,1}} 
   

   \institute{Max-Planck Institute for Radioastronomy,
              Auf dem H\"{u}gel, 53121\\
              \email{mtiwari@mpifr.mpg.de}
              \and European Southern Observatory, Alonso de C\'{o}rdova 3107,Vitacura Casilla 7630355, Santiago, Chile
              \and Instituto de Radioastronom\'ia Milim\'etrica, Avenida Divina Pastora 7, 18012 Granada, Spain}

   \date{Received September ....; accepted ...}

 \abstract
   {}
   {Hydrocarbons are ubiquitous in the interstellar medium, but their formation is still not well understood, depending on the physical environment they are found in. Messier 8 (M8) is host to one of the brightest \hii\ regions and photodissociation regions (PDRs) in our galaxy. With the observed C$_{2}$H and c-C$_{3}$H$_{2}$ data toward M8, we aim at obtaining their densities and abundances and to shed some light on their formation mechanism.}
   {Using the Atacama Pathfinder Experiment (APEX)~12~m, and the Institut de Radioastronomie Millim\'{e}trique (IRAM)~30~m telescopes, we performed a line survey toward Herschel 36 (Her 36), which is the main ionizing stellar system in M8, and an imaging survey within 1.3 $\times$ 1.3~pc around Her 36 of various transitions of C$_{2}$H and C$_{3}$H$_{2}$. We used both Local Thermodynamic Equilibrium (LTE) and non-LTE methods to determine the physical conditions of the emitting gas along with the column densities and abundances of the observed species, which we compared with (updated) gas phase photochemical PDR models. In order to examine the role of polycyclic aromatic hydrocarbons (PAHs) in the formation of small hydrocarbons and to investigate their association with the \hii\ region, the PDR and the molecular cloud, we compared archival Galactic Legacy Infrared Mid-Plane Survey Extraordinaire (GLIMPSE) 8~$\mu$m and the Spectral and Photometric Imaging Receiver (SPIRE)~250~$\mu$m continuum images with the C$_{2}$H emission maps.} 
   {We observed a total of three rotational transitions of C$_{2}$H with their hyperfine structure components and four rotational transitions of C$_{3}$H$_{2}$ with ortho and para symmetries toward the \hii\ region and the PDR of M8. Fragmentation of PAHs seems less likely to contribute to the formation of small hydrocarbons as the 8~$\mu$m emission does not follow the distribution of C$_{2}$H emission, which is more associated with the molecular cloud toward the north-west of Her 36. From the quantitative analysis, we obtained abundances of $\sim$~10$^{-8}$ and 10$^{-9}$ for C$_{2}$H and c-C$_{3}$H$_{2}$ respectively, and volume densities of the hydrocarbon emitting gas in the range $n(\rm H_2)$ $\sim$ 5 $\times$ 10$^{4}$--5 $\times$ 10$^{6}$~cm$^{-3}$.}
   {The observed column densities of C$_{2}$H and c-C$_3$H$_{2}$ are reproduced reasonably well by our PDR models. This supports the idea that in high-UV flux PDRs, gas phase chemistry is sufficient to explain hydrocarbon abundances.}

\authorrunning{M. Tiwari et al.}
   \maketitle
%
\section{Introduction}
Bright O and early B-type stars have strong ultraviolet (UV, h$\nu$ $>$ 13.6~eV) and far-ultraviolet (FUV, 6~eV $<$ h$\nu$ $<$ 13.6~eV) fields which give rise to bright \hii\ regions and photodissociation regions (PDRs). \hii\ regions comprise of hot ionized gas irradiated by strong UV radiations from nearby bright stars, while PDRs are at the interface of these \hii\ regions and cold molecular clouds shielded from the illuminating star \citep{1985ApJ...291..722T}. In PDRs, the heating is regulated by FUV photons which give rise to a rich hydrocarbon chemistry (eg. \citealt{2005A&A...435..885P}). Several studies on small hydrocarbons such as C$_{2}$H and C$_{3}$H$_{2}$ have been done in diffuse clouds \citep{2000A&A...358.1069L, 2011A&A...525A.116G}, massive star-forming regions \citep{2008ApJ...675L..33B}, planetary nebulae \citep{2017ApJ...850..123S}, dark clouds \citep{1997ApJ...486..862P}, and PDRs \citep{2004A&A...417..135T,2005A&A...435..885P,2015A&A...575A..82C,2015A&A...578A.124N}. These hydrocarbons play a key role in understanding the carbon chemistry taking place in both the gas phase and on grain surfaces. Several chemical pathways have been proposed for the formation of these hydrocarbons. It was found that in PDRs with low UV-flux ($G_0 <$ 100 in units of the \citet{1969BAN....20..177H} radiation field, where $G_0 = 1$ corresponds to a flux of $1.6\times 10^{-3}$~erg~cm$^{-2}$~s$^{-1}$ for FUV photons) such as the Horsehead nebula, the observed abundances of hydrocarbons were found to be higher than those predicted by gas-phase chemical PDR models by an order of magnitude \citep{2004A&A...417..135T}. This suggested that the current gas-phase chemistry does not adequately explain the observed high abundances of hydrocarbons, which led to the proposition that additional mechanisms may be responsible for their production. \citet{2003ApJ...584..316L}, \citet{2005A&A...435..885P} and \citet{2013A&A...552A..15M} suggest that one such mechanism is the fragmentation of polycyclic aromatic hydrocarbons (PAHs) due to far-UV radiation. While, in PDRs with a high UV-flux ($G_0 \sim$ 10$^{4}$--10$^{5}$ in Habing units) such as the Orion bar, the observed hydrocarbon abundances can be roughly explained by gas-phase chemistry. The molecular gas gets heated to higher temperatures in high UV flux PDRs facilitating new gas-phase formation routes, namely endothermic reactions, and reactions with activation energy barriers \citep{2015A&A...575A..82C}.\

Messier 8 (M8) is among the highest UV-flux ($G_0 \sim$ 10$^{5}$ in Habing units) \citep{refId0} PDRs in the Milky Way. It is located in the Sagittarius-Carina arm, near our line of sight toward the Galactic Center. Its distance from the sun has been estimated as $\sim$ 1.3~kpc, a value we adopt \citep{2004ApJ...608..781D,2006MNRAS.366..739A}; see the discussion in \citet{2008hsf2.book..533T}. The open young stellar cluster NGC6530, the \hii\ region NGC6523/33, and large quantities of molecular gas are associated with M8 \citep{2008hsf2.book..533T}. The brightest star in the open cluster NGC6530 is Herschel 36 (Her 36) at R.A.\,({\bf $\alpha$,} J2000) = 18$^{h}$03$^{m}$40$^{s}$.3 and Dec.\,({\bf $\delta$,} J2000) = $-$24$\degree$22$\arcmin$43$\arcsec$ \citep{1961PASP...73..206W}, which has three main resolved components: a close massive binary consisting of an O9 V and a B0.5 V star and a more distant companion O7.5 star (\citealt{2010ApJ...710L..30A} and \citealt{2014A&A...572L...1S}).\

An extensive survey of M8 at submillimeter, millimeter and far-infrared wavelengths was reported in \citet{refId0}, where we explored in detail the morphology of the area around Her 36 and determined the physical conditions governing this region. Observations were performed using several receivers of the Stratospheric Observatory for Infrared Astronomy (SOFIA, \citealt{2012ApJ...749L..17Y}), the Atacama Pathfinder EXperiment (APEX~12~m, \citealt{2006SPIE.6267E..14G}) and the Institut de Radioastronomie Millim\'{e}trique (IRAM)~30~m telescopes. M8 has a face-on geometry, where the cold dense molecular cloud lies in the background with Her 36 being still very close to the dense core of the cloud from which it originated. Her 36 is fueling the \hii\ region toward the east of it along with 9 Sagitarii (9 Sgr) (another binary with an O3.5 V and an O5-5.5V stars, \citealt{2012A&A...542A..95R}) and a foreground veil of warm PDR gas is receding away from Her 36 toward the observer (see the sketch in Fig.~15 of \citealt{refId0}). Using various CO transitions from $J$ = 1 $\to$ 0 to 16 $\to$ 15 at angular resolutions of $\sim$ 10$\arcsec$ (minimum) for CO $J =$ 6 $\to$ 5 to 30$\arcsec$ (maximum) for CO $J =$ 2 $\to$ 1 \citep[Table~1]{refId0}, we determined the average kinetic temperatures in the region ranging from 100--150~K and the H$_{2}$ densities ranging from 10$^{4}$--10$^{6}$~cm$^{-3}$, with an H$_{2}$ column density of $N(\rm H_{2})$ $\sim$ 3.8 $\times$ 10$^{22}$~cm$^{-2}$. 


\begin{table*}
\tiny
\centering
\begin{threeparttable}
\caption{Observed hydrocarbons toward M8.}
\begin{tabular}{c c . . c c c}
\hline\hline
\noalign{\smallskip}
 Species &Transition & \multicolumn{1}{c}{Frequency (GHz)} & \multicolumn{1}{c}{E$_{up}$/k (K)} &Instrument & Beam ($\arcsec$) & Critical density\tnote{a}~(cm$^{-3}$)\\
  \hline
 \noalign{\smallskip}
\multirow{18}{*}{C$_{2}$H} & $N$ = 1 $\to$ 0, $J$ = 3/2 $\to$ 1/2, $F$ = 1 $\to$ 1 & 87.2841& 4.2 & IRAM 30m/EMIR &30 & 2.5 $\times$ 10$^{5}$\\
 & $N$ = 1 $\to$ 0, $J$ = 3/2 $\to$ 1/2, $F$ = 2 $\to$ 1 & 87.3168& 4.2 &IRAM 30m/EMIR &30 &3.2 $\times$ 10$^{5}$\\
 & $N$ = 1 $\to$ 0, $J$ = 3/2 $\to$ 1/2, $F$ = 1 $\to$ 0 & 87.3285& 4.2 &IRAM 30m/EMIR &30 &9.5 $\times$ 10$^{4}$\\
 & $N$ = 1 $\to$ 0, $J$ = 1/2 $\to$ 1/2, $F$ = 1 $\to$ 1 & 87.4019& 4.2 &IRAM 30m/EMIR &30 &1.1 $\times$ 10$^{6}$\\
 & $N$ = 1 $\to$ 0, $J$ = 1/2 $\to$ 1/2, $F$ = 0 $\to$ 1 & 87.4071& 4.2 &IRAM 30m/EMIR &30  &1.8 $\times$ 10$^{5}$\\
 & $N$ = 1 $\to$ 0, $J$ = 1/2 $\to$ 1/2, $F$ = 1 $\to$ 0 & 87.4464& 4.2 &IRAM 30m/EMIR &30 &2.2 $\times$ 10$^{5}$\\
 
& $N$ = 3 $\to$ 2, $J$ = 7/2 $\to$ 5/2, $F$ = 3 $\to$ 3 & 261.9781& 25.1 & APEX/PI230 &26 &3.2 $\times$ 10$^{4}$\\
& $N$ = 3 $\to$ 2, $J$ = 7/2 $\to$ 5/2, $F$ = 4 $\to$ 3 & 262.0042& 25.1 & APEX/PI230 &26 &8.7 $\times$ 10$^{5}$\\
& $N$ = 3 $\to$ 2, $J$ = 7/2 $\to$ 5/2, $F$ = 3 $\to$ 2 & 262.0064& 25.1 & APEX/PI230 &26 &8.5 $\times$ 10$^{5}$\\
& $N$ = 3 $\to$ 2, $J$ = 5/2 $\to$ 3/2, $F$ = 3 $\to$ 2 & 262.0649& 25.1 & APEX/PI230 &26  &8.9 $\times$ 10$^{5}$\\
& $N$ = 3 $\to$ 2, $J$ = 5/2 $\to$ 3/2, $F$ = 2 $\to$ 1 & 262.0674& 25.1 & APEX/PI230 &26  &8.2 $\times$ 10$^{5}$\\
& $N$ = 3 $\to$ 2, $J$ = 5/2 $\to$ 3/2, $F$ = 2 $\to$ 2 & 262.0789& 25.1 & APEX/PI230 &26  &1.2 $\times$ 10$^{5}$\\
& $N$ = 3 $\to$ 2, $J$ = 5/2 $\to$ 5/2, $F$ = 3 $\to$ 3 & 262.2086& 25.1 & APEX/PI230 &26 &8.4 $\times$ 10$^{4}$\\
& $N$ = 3 $\to$ 2, $J$ = 5/2 $\to$ 5/2, $F$ = 2 $\to$ 2 & 262.2509& 25.1 & APEX/PI230 &26 &4.9 $\times$ 10$^{4}$\\

& $N$ = 5 $\to$ 4, $J$ = 11/2 $\to$ 9/2, $F$ = 6 $\to$ 5&436.661&62.9&APEX/FLASH$^{+}$ 460 &13  &2.9 $\times$ 10$^{6}$\\ 
& $N$ = 5 $\to$ 4, $J$ = 11/2 $\to$ 9/2, $F$ = 5 $\to$ 4&436.6618&62.9&APEX/FLASH$^{+}$ 460 &13 &2.9 $\times$ 10$^{6}$\\
& $N$ = 5 $\to$ 4, $J$ = 9/2 $\to$ 7/2, $F$ = 5 $\to$ 4&436.723&62.9&APEX/FLASH$^{+}$ 460 &13  &3.3 $\times$ 10$^{6}$\\
& $N$ = 5 $\to$ 4, $J$ = 9/2 $\to$ 7/2, $F$ = 4 $\to$ 3&436.724&62.9&APEX/FLASH$^{+}$ 460 &13 &3.2 $\times$ 10$^{6}$\\
\hline
\noalign{\smallskip}
\multirow{4}{*}{c-C$_{3}$H$_{2}$ ortho}& $J_{\rm K_{\rm a}, K_{\rm b}}$ = 2$_{1,2}$ $\to$ 1$_{0,2}$ &85.3388 & 6.4&IRAM 30m/EMIR & 30 &1.1 $\times$ 10$^{6}$\\
& $J_{\rm K_{\rm a}, K_{\rm b}}$ = 6$_{3,4}$ $\to$ 5$_{2,3}$ &285.7956 &54.7 &APEX/FLASH$^{+}$ 345 &19  &8.4 $\times$ 10$^{6}$\\
& $J_{\rm K_{\rm a}, K_{\rm b}}$ = 7$_{1,6}$ $\to$ 6$_{2,5}$ &284.998 &61.2 &APEX/FLASH$^{+}$ 345 &19 &8.9 $\times$ 10$^{6}$\\
& $J_{\rm K_{\rm a}, K_{\rm b}}$ = 8$_{1,8}$ $\to$ 7$_{0,7}$ &284.8052 &64.3 &APEX/FLASH$^{+}$ 345 &19 &1.0 $\times$ 10$^{7}$\\
\hline
\noalign{\smallskip}
\multirow{3}{*}{c-C$_{3}$H$_{2}$ para}& $J_{\rm K_{\rm a}, K_{\rm b}}$ = 2$_{0,2}$ $\to$ 1$_{1,1}$ &82.0935 &6.4 &IRAM 30m/EMIR & 30 &9.9 $\times$ 10$^{5}$\\
& $J_{\rm K_{\rm a}, K_{\rm b}}$ = 7$_{2,6}$ $\to$ 6$_{1,5}$ &284.9993 & 61.2&APEX/FLASH$^{+}$ 345 & 19 &1.0 $\times$ 10$^{7}$\\
& $J_{\rm K_{\rm a}, K_{\rm b}}$ = 8$_{0,8}$ $\to$ 7$_{1,7}$ &284.8052 & 64.3&APEX/FLASH$^{+}$ 345 & 19 &1.0 $\times$ 10$^{7}$\\
 \hline
\end{tabular}

\begin{tablenotes}
\item[a]\small Critical densities are calculated using collision rates available at $T_{\rm kin}$ = 100~K for C$_{2}$H and at $T_{\rm kin}$ = 120~K for c-C$_{3}$H$_{2}$ from \citet{2012MNRAS.421.1891S} and \citet{2000A&AS..142..113C} respectively.
\end{tablenotes}

 \label{appendix a}
 \end{threeparttable}
\end{table*} 

Several studies have addressed M8 in the X-ray, optical and IR regimes (\citealt{1995ApJ...445L.153S,2004ApJ...608..781D,2006MNRAS.366..739A,2006ApJ...649..299G,2017A&A...604A.135D}). Interestingly, \citet{2013ApJ...773...41D} found anomalously broad diffuse interstellar bands (DIBs) at 5780.5, 5797.1, 6196.0 and 6613.6~$\AA$ along with CH$^{+}$ and CH in absorption along the line of sight to Her 36. A bright IR source Her 36 SE \citep{2006ApJ...649..299G} lying 0.25$\arcsec$ south-east of Her 36, is responsible for radiative excitation of CH$^{+}$ and CH. The broadening of DIBs has been attributed to radiative pumping of closely spaced high-$J$ rotational levels of small polar molecules (\citealt{2013ApJ...773...41D,2014ApJ...793...68O,2014IAUS..297...89Y}).\

With the goal to constrain the physical conditions of the gas responsible for the emission of small hydrocarbons and to verify if gas phase chemistry alone can explain the observed abundance of small hydrocarbons in M8, we here present an inventory of small hydrocarbons found toward M8 in a comprehensive survey including both observations toward the bright stellar system Her 36 and On-The-Fly (OTF) maps in a region of 4$\arcmin$ $\times$ 4$\arcmin$ (corresponding to 1.5 $\times$ 1.5~pc) around Her 36 at millimeter- and submillimeter wavelengths. The organization of the paper is as follows. In Sect.~2, we summarize the observations done using the PI230 and FLASH$^{+}$ receivers of the APEX\footnote{This publication is based on data acquired with the Atacama Pathfinder EXperiment (APEX). APEX is a collaboration between the Max-Planck-Institut f{\"u}r Radioastronomie, the European Southern Observatory, and the Onsala Space Observatory.} 12~m telescope and the EMIR receiver of the IRAM\footnote{Based on observations carried out with the IRAM~30~m telescope. IRAM is supported by INSU/CNRS, the MPG (Germany), and IGN (Spain)}~30m telescope. In Sect.~3, we present the spectra and the velocity integrated maps of the observed transitions of C$_{2}$H and c-C$_{3}$H$_{2}$. The quantitative analysis of the data is described in Sect.~4. In Sect.~5, we discuss the results and the main conclusions of this work are summarized in Sect.~6.      
     
\section{Observations}
\subsection{The APEX data}
\begin{figure*}[t]
  \centering
  \subfigure{\includegraphics[width=160mm]{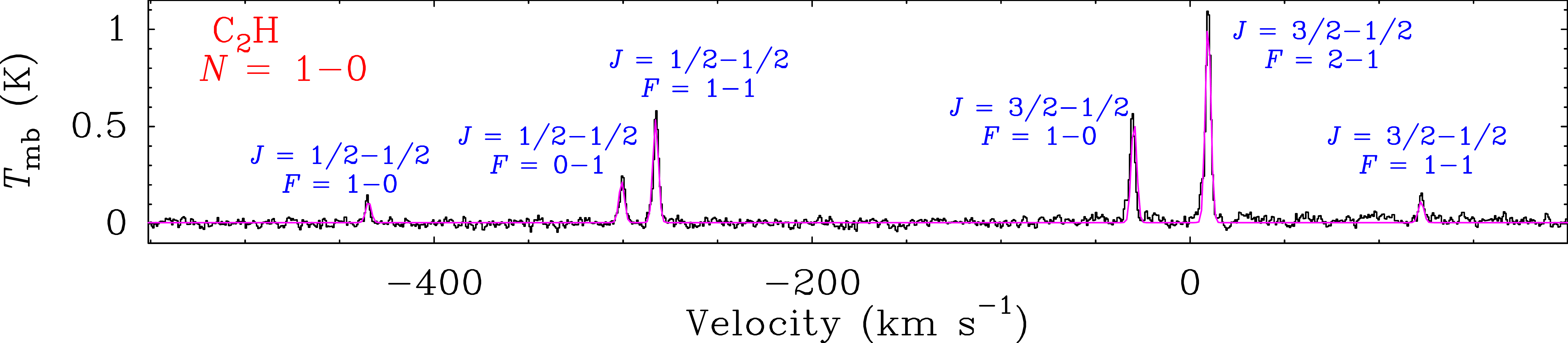}}
 \subfigure{\includegraphics[width=160mm]{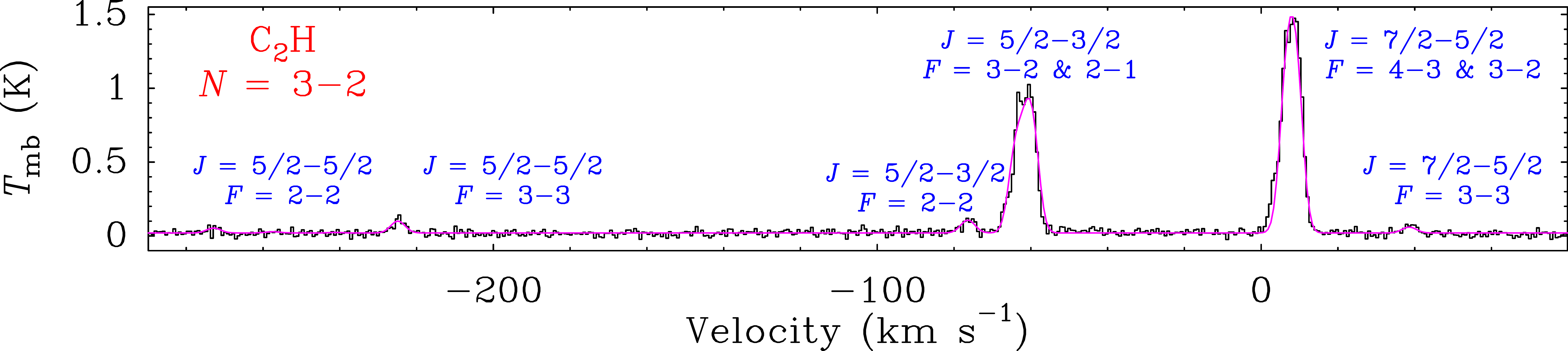}}
  \subfigure{\includegraphics[width=160mm]{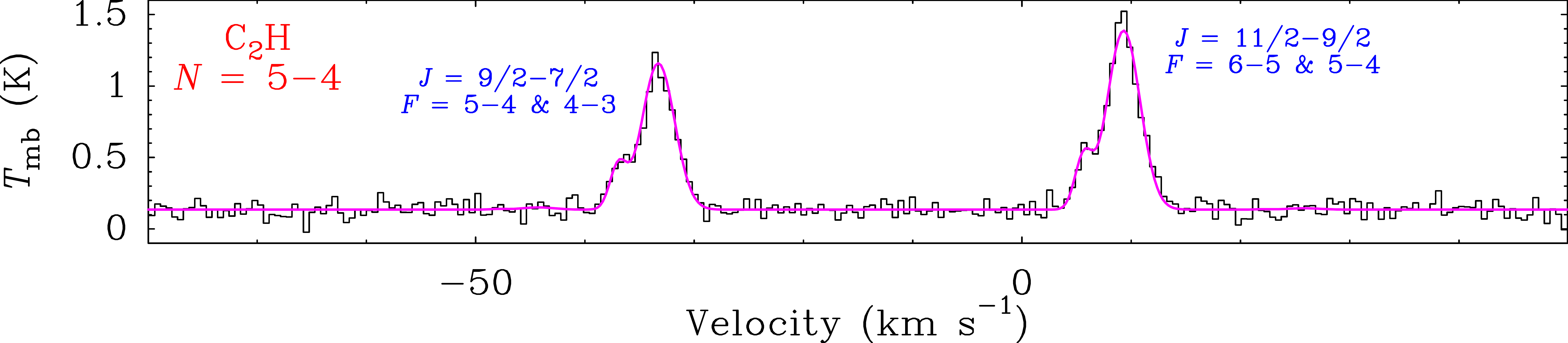}}

  \caption{Observed hfs components of the N = 1 $\to$ 0, 3 $\to$ 2, 5 $\to$ 4 rotational transitions of C$_{2}$H toward Her 36 at R.A. 18$^{h}$03$^{m}$40$^{s}$.3; Dec. $-$24$\degree$22$\arcmin$43$\arcsec$ (J2000). For the $N$ = 1 $\to$ 0 transition all 6 hfs are fully resolved, while for the $N$ = 3 $\to$ 2 and 5 $\to$ 4 transitions, some  hfs lines overlap with each other. The $N$ = 5 $\to$ 4 shows an additional spectrally resolved velocity component (2--8~km~s$^{-1}$) compared to other transitions. Only the detected lines are indicated by their quantum numbers $J$ and $F$.}
 \label{all_obs} 
\end{figure*}

Observations of the $N$ = 3 $\to$ 2 and 5 $\to$ 4 transitions of C$_{2}$H and $J_{\rm K_{\rm a}, K_{\rm b}}$ = 6$_{3,4}$ $\to$ 5$_{2,3}$, 7$_{1,6}$ $\to$ 6$_{2,5}$ and 8$_{1,8}$ $\to$ 7$_{0,7}$ ortho; and 7$_{2,6}$ $\to$ 6$_{1,5}$ and 8$_{0,8}$ $\to$ 7$_{1,7}$ para transitions of c-C$_{3}$H$_{2}$ were performed with the APEX~12~m submillimeter telescope during 2015 June--August and 2016 July and September. As shown in Table~\ref{all_obs}, we used the following receivers: PI230 with a velocity resolution of 0.07~km~s$^{-1}$, to map the $N$ = 3 $\to$ 2 of C$_{2}$H, which is split into 8 hyperfine structure (hfs) components, FLASH$^{+}$ in the 460 GHz band with a velocity resolution of 0.05~km~s$^{-1}$, to integrate deeply on the $N$ = 5 $\to$ 4 of C$_{2}$H, (4 hfs components) toward Her 36 and FLASH$^{+}$ in the 345 GHz band with a velocity resolution of 0.04~km~s$^{-1}$, to integrate deeply on the $J_{\rm K_{\rm a}, K_{\rm b}}$ = 6$_{3,4}$ $\to$ 5$_{2,3}$, 7$_{1,6}$ $\to$ 6$_{2,5}$ and 8$_{1,8}$ $\to$ 7$_{0,7}$ ortho; and 7$_{2,6}$ $\to$ 6$_{1,5}$ and 8$_{0,8}$ $\to$ 7$_{1,7}$ para transitions of c-C$_{3}$H$_{2}$ toward Her 36. To increase the signal-to-noise ratio, the data was later smoothed from a velocity resolution of 0.07~km~s$^{-1}$ to 0.7~km~s$^{-1}$, 0.05~km~s$^{-1}$ to 0.5~km~s$^{-1}$ and 0.04~km~s$^{-1}$ to 0.4~km~s$^{-1}$. \ 

The map was observed in on-the-fly (OTF) total power mode centered on  R.A. = 18$^{h}$03$^{m}$40$^{s}$.3 and Dec. = $-$24$\degree$23$\arcmin$12$\arcsec$(J2000) which corresponds to the position of Her 36. It has a size $\sim$ 240$\arcsec$ $\times$ 240$\arcsec$. Deep integrations were pointed at Her 36 for about $\sim$ 3~minutes while, we integrated 0.7~s per dump for the map. An offset position relative to the center at (30$\arcmin$, --30$\arcmin$) was chosen for reference, similar to the previous observations done in \citet{refId0}. The pointing accuracy ($<$ 3$\arcsec$) was maintained by pointing checks on bright sources such as Mars and R Dor with the receivers tuned to CO lines every 1--1.5~hrs. A forward efficiency $\eta_{\rm f}$ = 0.95 was used for all receivers, and the beam coupling efficiencies $\eta_{\rm c}$ = 0.62, 0.69 and 0.56 were used for the PI230, FLASH\textsuperscript{+}340 and FLASH\textsuperscript{+}460 receivers, respectively.\ 

\subsection{The IRAM~30~m data}
Observations of the $N$ = 1 $\to$ 0 transition of C$_{2}$H; and $J_{\rm K_{\rm a}, K_{\rm b}}$ = 2$_{1,2}$ $\to$ 1$_{0,2}$ ortho and 2$_{0,2}$ $\to$ 1$_{1,1}$ transitions of c-C$_{3}$H$_{2}$ were performed with the IRAM~30~m telescope in August 2016. Most of the 3~mm range was observed using the EMIR receivers \citep{2012A&A...538A..89C} with a velocity resolution of 0.65~km~s$^{-1}$. Six hfs structure lines of C$_{2}$H $N$ = 1 $\to$ 0 transition were mapped and deep integration pointed observations $J_{\rm K_{\rm a}, K_{\rm b}}$ = 2$_{1,2}$ $\to$ 1$_{0,1}$ ortho and 2$_{0,2}$ $\to$ 1$_{1,1}$ para transitions of c-C$_{3}$H$_{2}$ were performed toward Her 36.\

\begin{figure}[htp]
  \centering
  \subfigure{\includegraphics[width=41mm]{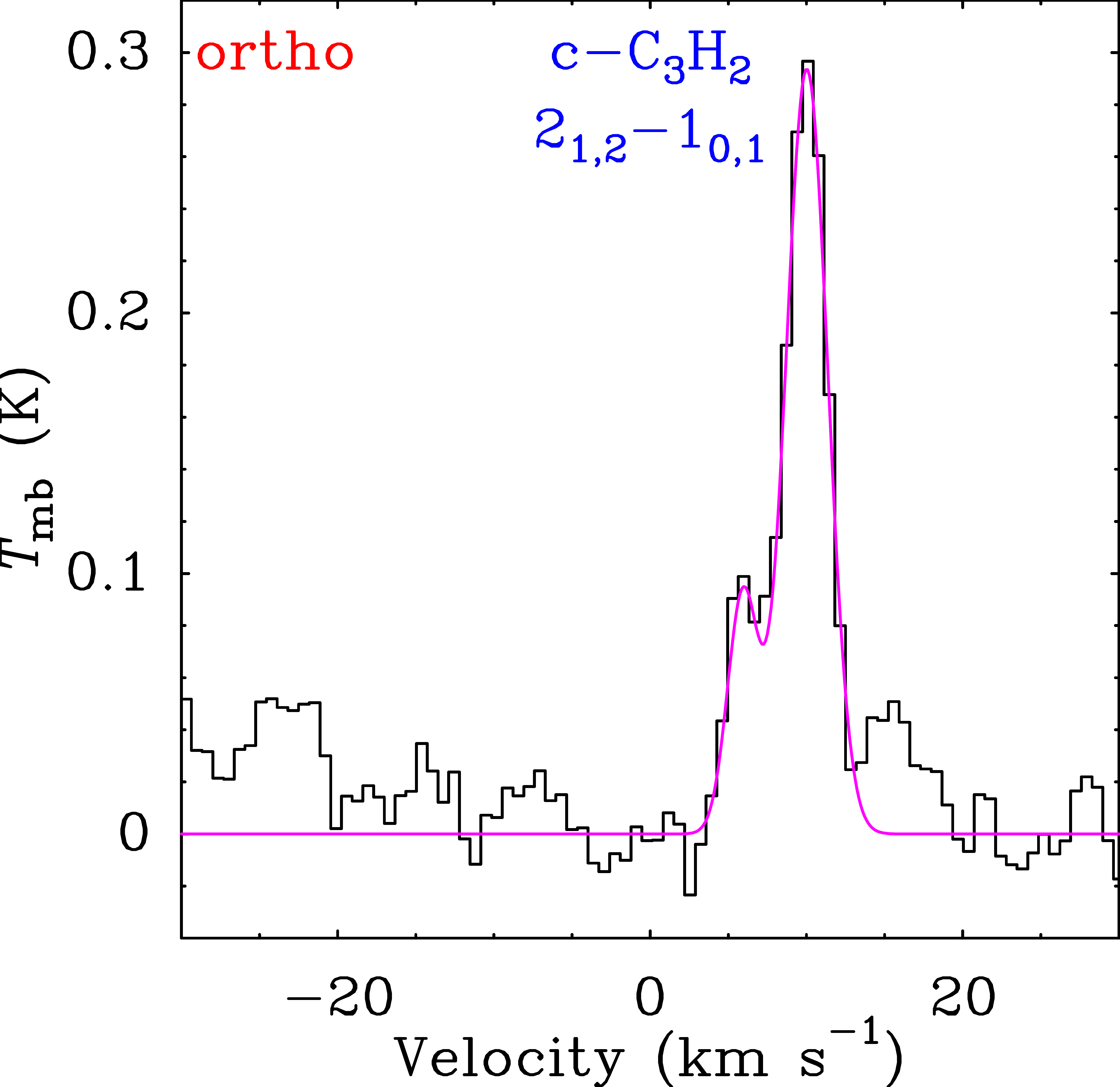}}\qquad
 \subfigure{\includegraphics[width=41mm]{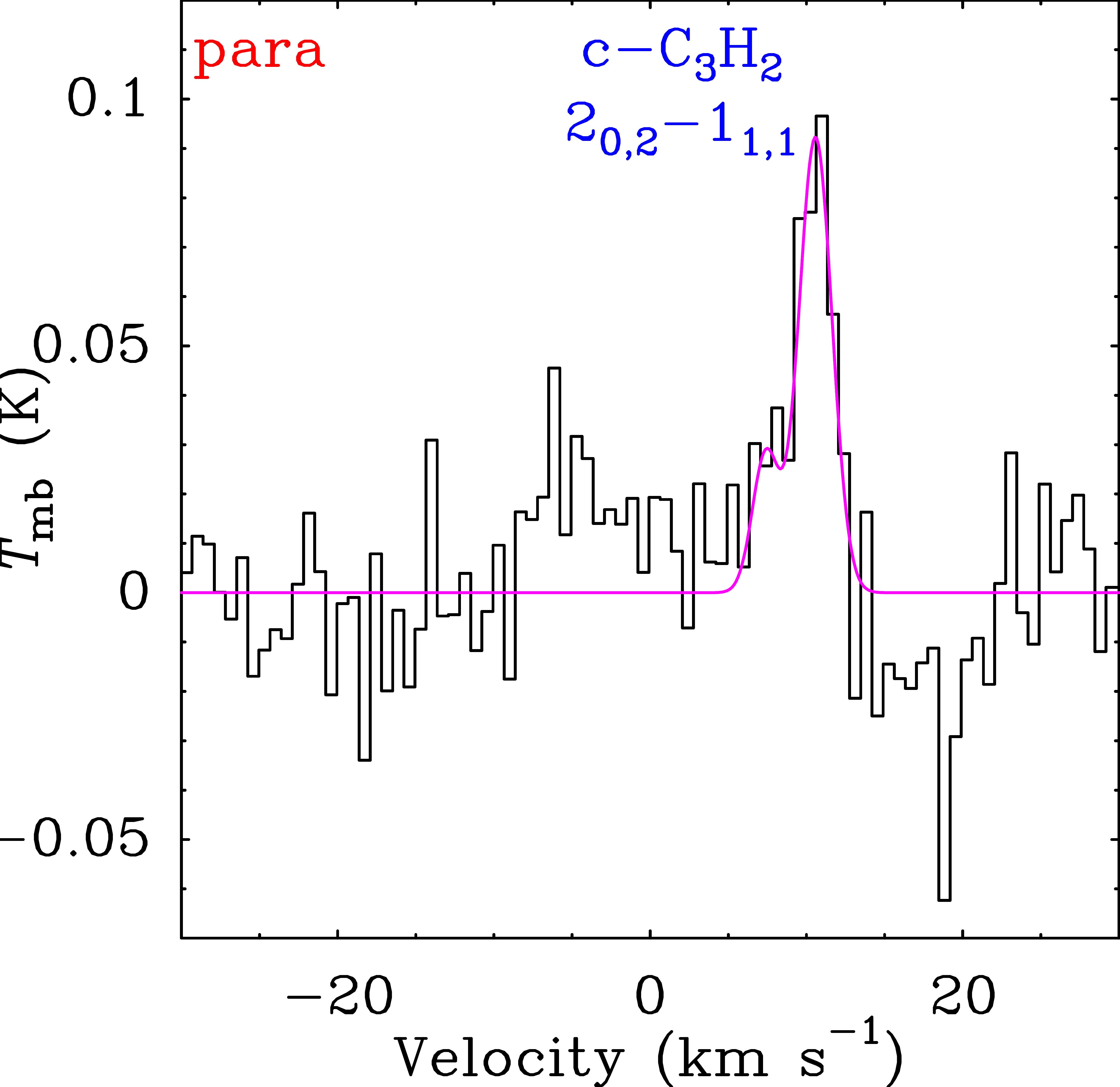}}
 \subfigure{\includegraphics[width=41mm]{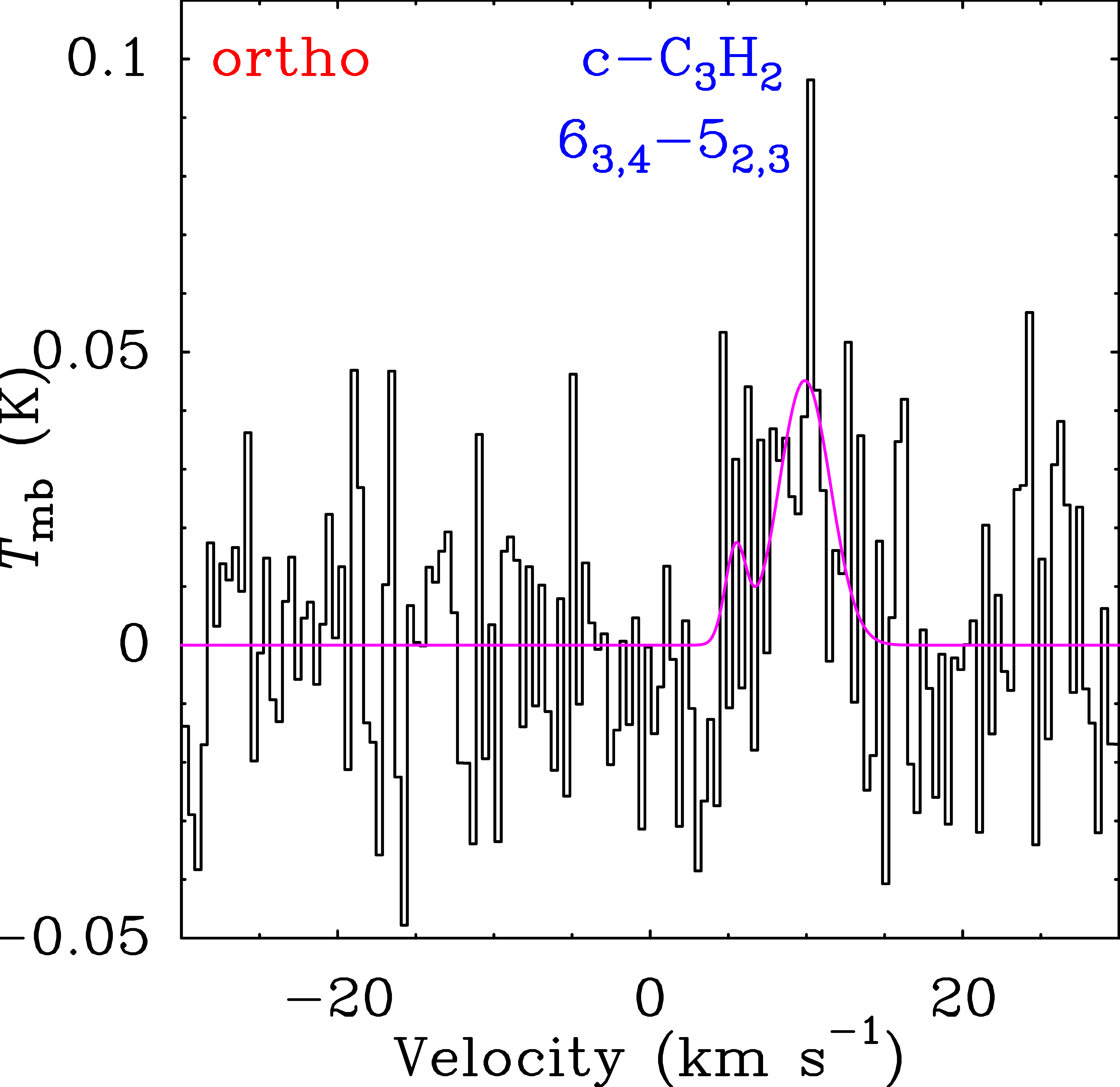}}\qquad
  \subfigure{\includegraphics[width=41mm]{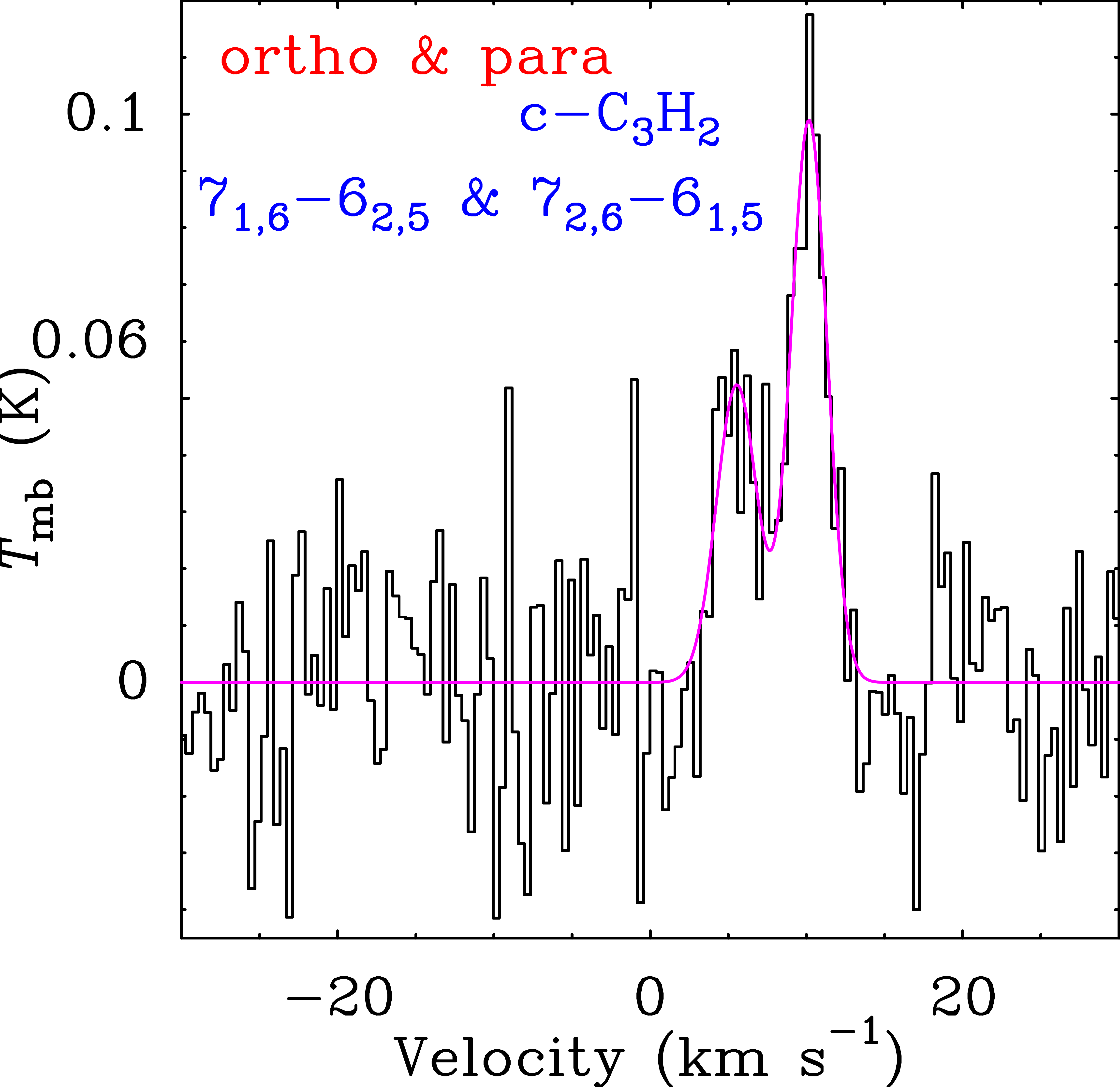}}
\subfigure{\includegraphics[width=41mm]{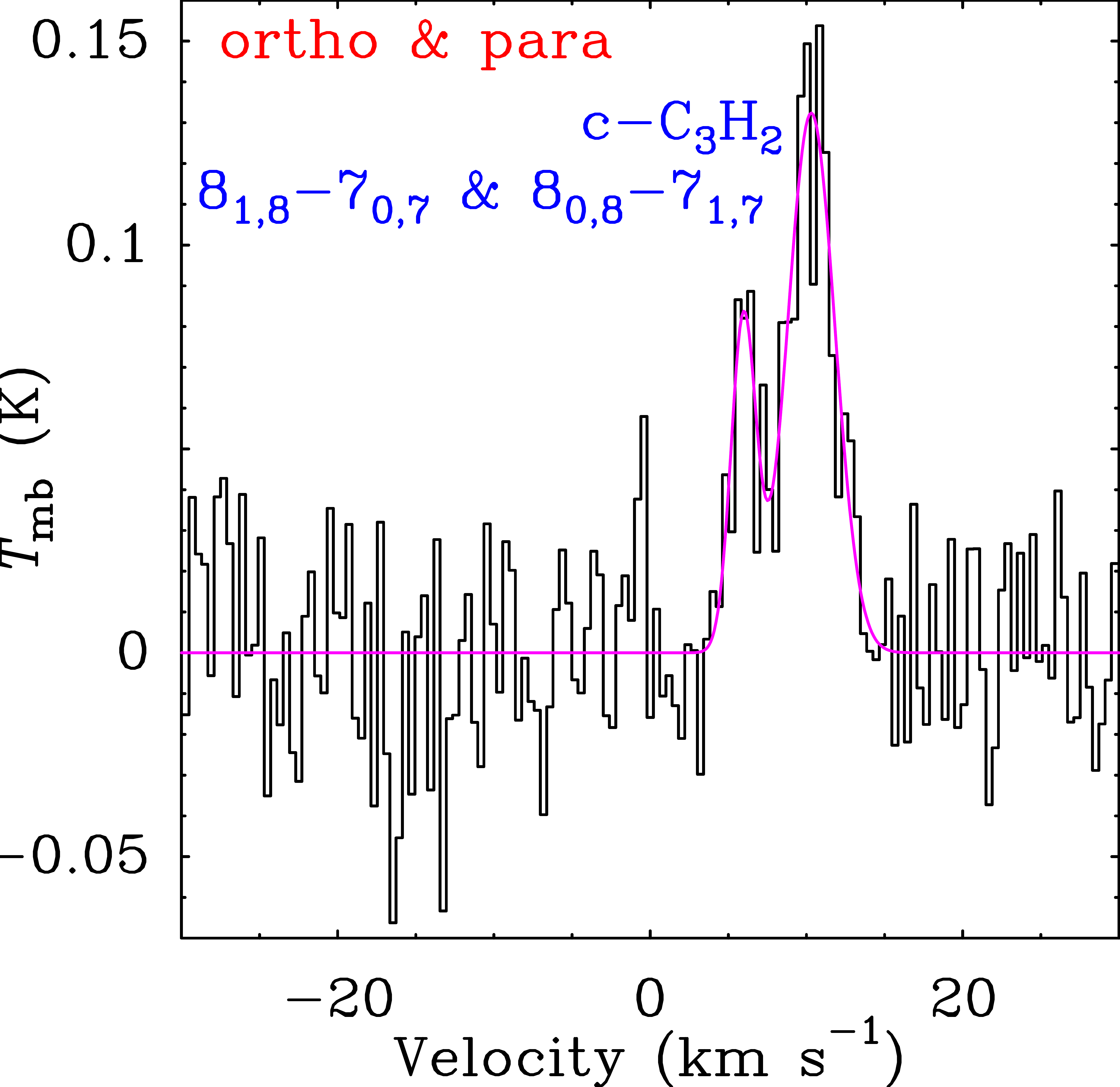}}

  \caption{Observations of spectrally resolved $J_{\rm K_{\rm a}, K_{\rm b}}$ = 2$_{1,2}$ $\to$ 1$_{0,1}$ and 6$_{3,4}$ $\to$ 5$_{2,3}$ ortho and $J_{\rm K_{\rm a}, K_{\rm b}}$ = 2$_{0,2}$ $\to$ 1$_{1,1}$ para transitions of c-C$_3$H$_2$ along with the spectrally blended 7$_{1,6}$ $\to$ 6$_{2,5}$ and 8$_{1,8}$ $\to$ 7$_{0,7}$ ortho; and 7$_{2,6}$ $\to$ 6$_{1,5}$ and 8$_{0,8}$ $\to$ 7$_{1,7}$ para transitions toward Her 36 at R.A. 18$^{h}$03$^{m}$40$^{s}$.3; Dec. $-$24$\degree$22$\arcmin$43$\arcsec$ (J2000).}
  
\end{figure}

Similar to APEX observations, the map was observed in OTF total power mode centered on Her 36 and has a size of 240$\arcsec$ $\times$ 240$\arcsec$. Each sub scan lasted 25~s and the integration time on the off-source reference position was 5~s. The offset position relative to the center at (30$\arcmin$, -30$\arcmin$) was similar to that used for APEX observations and the pointing accuracy ($<$ 3$\arcsec$) was maintained by pointing at the bright calibrator 1757-240 every 1$-$1.5~hrs. A forward efficiency $\eta_{\rm f}$ = 0.95 and the beam coupling efficiency $\eta_{\rm c}$ = 0.69 were adopted for EMIR receivers. These values were taken from the (2015) commissioning report\footnote{www.iram.es/IRAMES/mainWIKI/IRAM30mEfficiencies}.\ 

All data reduction employed the CLASS and GREG softwares that are a part of the GILDAS\footnote{www.iram.fr/IRAMFR/GILDAS/} software package and all observations are summarized in Table~\ref{all_obs}.

\begin{table}[ht]
\tiny
\centering
\begin{threeparttable}
\caption{Hfs fit parameters of the observed low and high velocity components of C$_{2}$H.}

\begin{tabular}{c c c c c}
 \hline\hline
 \noalign{\smallskip}
 Transition & $v$\tnote{a} (km s$^{-1}$) & $\Delta$~$v$\tnote{a} (km s$^{-1}$) & $T_{\rm mb}$\tnote{a} (K) & $\tau$\tnote{a} \\
 \hline
  \multicolumn{5}{c}{Low velocity component}\\
 \hline
   \noalign{\smallskip}

 $N$ = 5 $\to$ 4 & 5.9 (0.524) & 1.63 (1.74) &  0.9 (0.053) & 0.78 (0.1)\\
 
 \hline
 \multicolumn{5}{c}{High velocity component}\\
  \hline
   \noalign{\smallskip}
  
  $N$ = 1 $\to$ 0 & 9.5 (0.0025) & 3.43 (0.014) & 2.5 (0.018)  & $\sim$~0.1\tnote{b} \\
  $N$ = 3 $\to$ 2 & 8.8 (0.013)  & 4.52 (0.046) & 4.5 (0.031) & 1.74 (0.102) \\
  $N$ = 5 $\to$ 4 & 9.6 (0.524) & 3.45 (1.74) &  2.6 (0.053) & $\sim$~0.1\tnote{b}\\ 
   
  \hline
 \noalign{\smallskip}
 \end{tabular}

\begin{tablenotes}
\item[a]\small The values in the parenthesis indicate the errors from the fit.
\item[b]\small 0.1 is the lower limit value of $\tau$ calculated by GILDAS for optically thin emission.
\end{tablenotes}

 \label{line_par_cch}
 \end{threeparttable}
\end{table}

\section{Results}
\subsection{Ethynyl: C$_{2}$H}

The ethynyl radical (C$_{2}$H) was first detected in the ISM by \citet{1974ApJ...193L.115T}. It is a linear molecule with spin rotation and hyperfine structure. The energy levels are designated as $N$, $J$ and $F$. The coupling between the rotational angular momentum $N$ and the unpaired electron spin $S$, causes spin doubling ($J$ = $N$ + $S$), while the coupling of angular momentum $J$ and spin of the hydrogen nucleus $I$, results in hfs ($F$ = $J$ + $I$) \citep{2015A&A...575A..82C}. The electric dipole selection rules allow hfs splitting to occur only in specific quantum levels. We were able to identify a total of 18 hfs components of C$_{2}$H and the line parameters derived from hfs fitting. The C$_{2}$H $N$ = 1 $\rightarrow$ 0 and 3 $\rightarrow$ 2 transitions have a single velocity component, while C$_{2}$H $N$ = 5 $\rightarrow$ 4 also have a spectrally resolved low velocity (2--8~km~s$^{-1}$) component, in addition to a high velocity (8--15 km~s$^{-1}$) component. This low velocity component is also seen in the CO, \cii\ and \ci\ spectra and corresponds to the warm foreground veil receding away from Her 36 toward the observer \citep{refId0}. The hfs fit parameters of the observed low and high velocity components of C$_{2}$H are given in Table~\ref{line_par_cch}.\

The hfs line fitting results for all the transitions are shown in Fig.~1. The lowest energy rotational transition $N$ = 1 $\rightarrow$ 0 is split into 6 observable hfs levels that are well separated in frequency. The observed relative intensities are listed in Table~\ref{rel_int_cch} and it can be seen except for the lowest intensity lines, all match well with the expected relative intrinsic intensities as obtained from the Cologne Database for Molecular Spectroscopy\footnote{http://www.astro.uni-koeln.de/cdms/} (CDMS). Hence, the lines are optically thin with no hfs emission anomalies and consistent with an optical depth of $\tau \sim$ 0.1. For the $N$ = 3 $\rightarrow$ 2 rotational transition, we identified 8 hfs components. Among them, some hfs features are not completely resolved and hence, some lines overlap. The $J$ = 7/2 $\rightarrow$ 5/2, $F$ = 4 $\rightarrow 3$ and $F$ = 3 $\rightarrow$ 2 overlap; and the $J$ = 5/2 $\rightarrow$ 3/2, $F$ = 3 $\rightarrow 2$ and $F$ = 2 $\rightarrow$ 1 overlap. The observed relative line intensities come out to be different from the expected relative intensities except for the higher transitions as shown in Table~\ref{rel_int_cch}. This suggests optical depth effects and indeed an optical depth of $\tau \sim$ 1.74 was calculated. For the $N$ = 5 $\rightarrow$ 4 rotational transition, we identified 4 hfs components where the lines with $J$ = 9/2 $\rightarrow$ 7/2, $F$ = 5 $\rightarrow 4$ and $F$ = 4 $\rightarrow$ 3; and the lines $J$ = 11/2 $\rightarrow$ 9/2, $F$ = 6 $\rightarrow 5$ and $F$ = 5 $\rightarrow$ 4 overlap. The observed relative intensities match well with the expected relative intrinsic intensities and are consistent within optical depth $\tau \sim$ 0.1.\\  

\begin{figure}[htp]
  \centering
  \subfigure{\includegraphics[width=80mm]{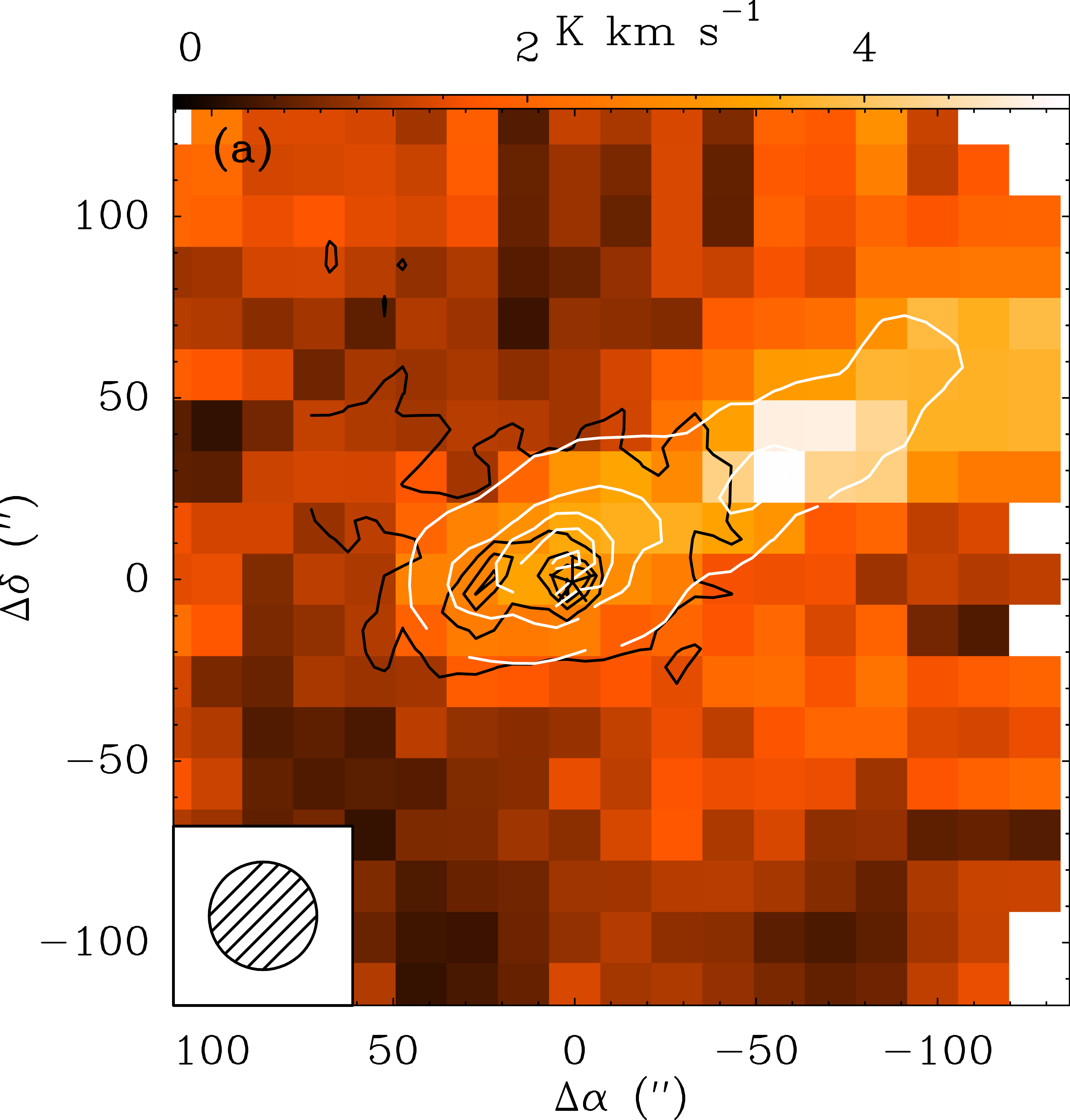}}\quad
 \subfigure{\includegraphics[width=80mm]{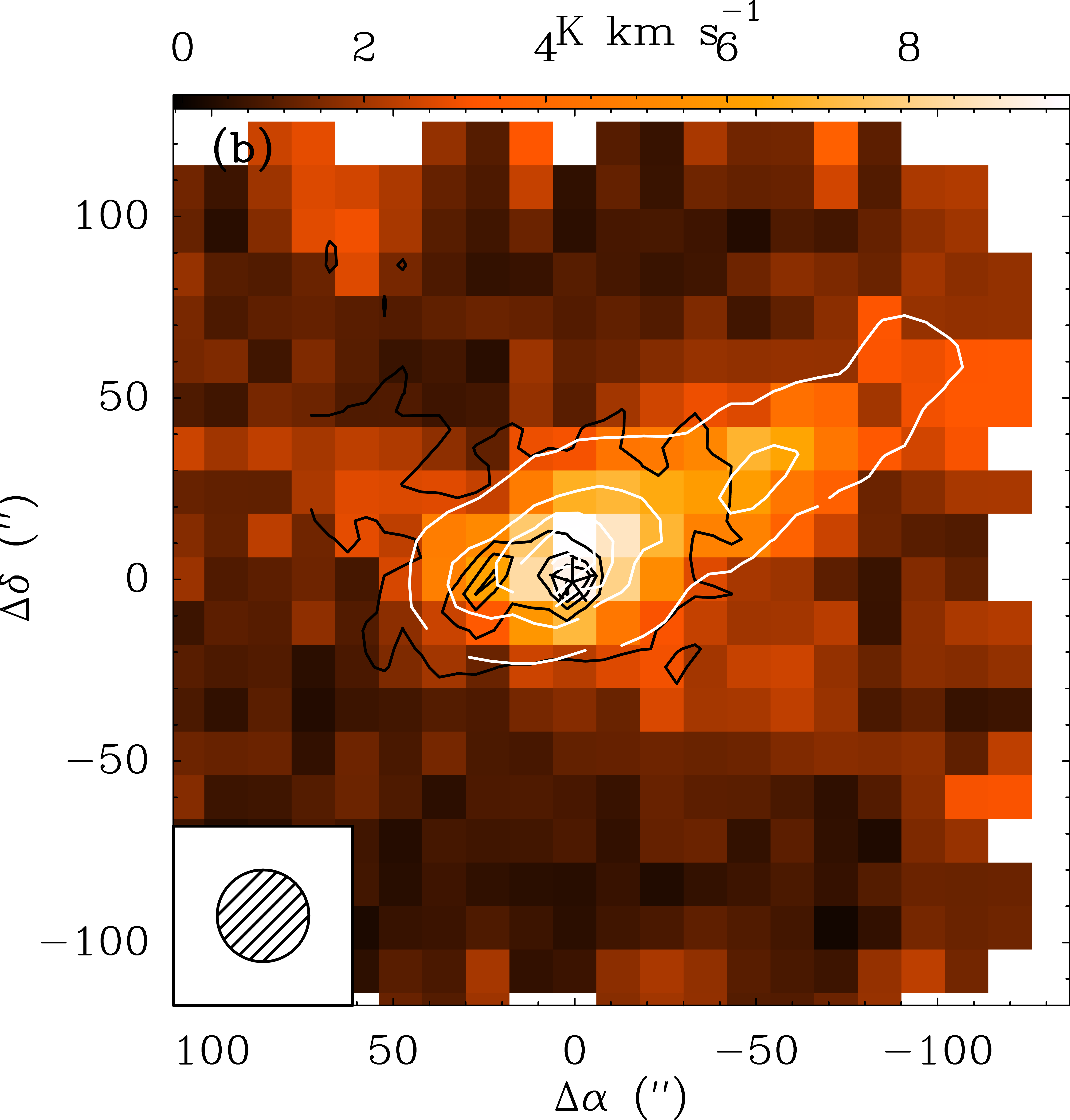}}

  \caption{Colour maps of the velocity integrated intensity of the: (a, upper panel) $N$ = 1 $\to$ 0 and (b, lower panel) 3 $\to$ 2 transitions of C$_{2}$H toward Her 36 which is the central position ($\Delta\alpha$ = 0, $\Delta\delta$ = 0) at R.A.(J2000) = 18\textsuperscript{h}03\textsuperscript{m}40.3\textsuperscript{s} and Dec.(J2000) = --24\degree22$\arcmin$43$\arcsec$, marked with an asterisk. Both maps are plotted using original beam sizes shown in the bottom left of each map. These maps are overlaid with contours of GLIMPSE 8~$\mu$m continuum emission in black, with a beam size of 0.6$\arcsec$ and with contours of SPIRE 250~$\mu$m continuum emission in white, with a beam size of 17.9$\arcsec$. For GLIMPSE 8~$\mu$m continuum emission, the contour levels are 10$\%$ to 100$\%$ in steps of 20$\%$ of the peak emission and for SPIRE 250~$\mu$m continuum emission, the contour levels are 10$\%$ to 100$\%$ in steps of 10$\%$ of the peak emission.}

 \label{vel_int_maps} 
\end{figure}

\begin{table}[ht]
\tiny
\centering
\begin{threeparttable}
\caption{C$_3$H$_2$ line parameters as calculated from two component gaussian fit using GILDAS.}

\begin{tabular}{c c c c c }
 \hline\hline
 \noalign{\smallskip}
 Transition & Symmetry &$v$\tnote{a} (km s$^{-1}$) & $\Delta$~$v$\tnote{a} (km s$^{-1}$) & $T_{\rm peak}$\tnote{a} \textbf{(mK)}  \\
 \hline
  \multicolumn{5}{c}{Low velocity component}\\
 \hline
   \noalign{\smallskip}

 2$_{1,2}$ $\to$ 1$_{0,1}$ & ortho & 6.0 (0.22) & 2.3 (0.43) & 90 (18)\\
 2$_{0,2}$ $\to$ 1$_{1,1}$ & para & 7.2 (0.62) & 2.0 (0) & 40 (17)\\
 6$_{3,4}$ $\to$ 5$_{2,3}$ & ortho & 6.3 (1.0) & 2.7 (1.3) & 20 (23)\\
 7$_{1,6}$ $\to$ 6$_{2,5}$ & ortho & 5.5 (0.3) & 3.1 (0.7) & 50 (13)\\
 7$_{2,6}$ $\to$ 6$_{1,5}$ & para & 5.5 (0.3) & 3.1 (0.7) & 50 (13)\\
 8$_{1,8}$ $\to$ 7$_{0,7}$ & ortho & 6.0 (0.2) & 1.9 (0.5) & 80 (20)\\
8$_{0,8}$ $\to$ 7$_{1,7}$ & para & 6.0 (0.2) & 1.9 (0.5) & 80 (20)\\

 \hline
 \multicolumn{5}{c}{High velocity component}\\
  \hline
   \noalign{\smallskip}
  2$_{1,2}$ $\to$ 1$_{0,1}$ & ortho & 10 (0.07) & 3.17 (0.2) & 300 (18)\\
  2$_{0,2}$ $\to$ 1$_{1,1}$ & para & 10.5 (0.2) & 2.6 (0.4) & 100 (17)\\
  6$_{3,4}$ $\to$ 5$_{2,3}$ & ortho & 10.1 (0.3) & 3.3 (1.2) & 50 (23)\\  
  7$_{1,6}$ $\to$ 6$_{2,5}$ & ortho & 10.2 (0.15) & 2.7 (0.4) & 100 (13)\\
  7$_{2,6}$ $\to$ 6$_{1,5}$ & para & 10.2 (0.15) & 2.7 (0.4) & 100 (13)\\
  8$_{1,8}$ $\to$ 7$_{0,7}$ & ortho & 10.3 (0.15) &3.6 (0.4) & 130 (20)\\
  8$_{0,8}$ $\to$ 7$_{1,7}$ & para & 10.3 (0.15) &3.6 (0.4) & 130 (20)\\
  \hline
 \noalign{\smallskip}
 
  \end{tabular}
  
  \begin{tablenotes}
\item[a]\small The values in the parenthesis indicate the errors from the fit.
\end{tablenotes}
  
 \label{line_par_c3h2}
 \end{threeparttable}
\end{table}

\subsection{Cyclopropenylidene: c-C$_3H_2$} 
Cyclopropenylidene, c-C$_{3}$H$_{2}$, was first discovered in the ISM by \citet{1985ApJ...298L..61M} and \citet{1985ApJ...294L..49T}. It is a three-membered carbon ring with C$_{\rm 2v}$ symmetry (see \citealt[Fig.~1]{2012ApJS..200....1S}), with a large dipole moment of 3.4 Debye making it highly polar. It has ortho and para symmetries owing to the two out of plane hydrogen atoms, which are at equidistant from the C atoms. It is an oblate asymmetric top with b-type rotational transitions where the main selection rules are $\Delta$$K_{\rm a}$ and $\Delta$$K_{\rm c}$ = $\pm$1. $K_{\rm a}$ + $K_{\rm c}$ = odd/even values correspond to the ortho/para levels (for more extensive description see \citealt{2015A&A...575A..82C}).\

We observed 2 completely resolved ortho transitions $J_{\rm K_{\rm a}, K_{\rm b}}$ = 2$_{1,2}$ $\to$ 1$_{0,1}$ and 6$_{3,4}$ $\to$ 5$_{2,3}$ and 1 para transition $J_{\rm K_{\rm a}, K_{\rm b}}$ = 2$_{0,2}$ $\to$ 1$_{1,1}$. In addition, we observed $J_{\rm K_{\rm a}, K_{\rm b}}$ = 7$_{1,6}$ $\to$ 6$_{2,5}$ and 8$_{1,8}$ $\to$ 7$_{0,7}$ ortho; and 7$_{2,6}$ $\to$ 6$_{1,5}$ and 8$_{0,8}$ $\to$ 7$_{1,7}$ para transitions where ortho and para lines are blended. Similar to the $N$ = 5 $\rightarrow$ 4 transition of C$_{2}$H, all observed transitions of c-C$_{3}$H$_{2}$ have a spectrally resolved low velocity component (centered at $\sim$ 6~km~s$^{-1}$)  as can be seen in Fig.~2. The line parameters are derived from a two component gaussian fit to the observed spectra as mentioned in Table~\ref{line_par_c3h2}.      
     
\subsection{Spatial distribution of C$_{2}$H and comparison with ancillary data}
Figure~3 shows velocity integrated intensity maps of the brightest transitions of C$_{2}$H $N$ = 1 $\to$ 0, $J$ = 3/2 $\to$ 1/2, $F$ = 2 $\to$ 1 at 87.316~GHz with an rms noise of 0.3~K~km~s$^{-1}$ and $N$ = 3 $\to$ 2 $J$ = 7/2 $\to$ 5/2, $F$ = 4 $\to$ 3 at 262.004~GHz with an rms noise of 0.44~K~km~s$^{-1}$. The $N$ = 1 $\to$ 0 transition is bright toward the north-west of Her 36 and peaks (in projection) deep into the molecular cloud at ($\Delta\alpha$ = --56$\arcsec$, $\Delta\delta$ = 28$\arcsec$). This position is very close to the secondary ATLASGAL~870~$\mu$m dust continuum peak toward this region at ($\Delta\alpha$ = --53$\arcsec$, $\Delta\delta$ = 23$\arcsec$) \citep[Fig.~7~(a)]{refId0} which is probing the cold molecular cloud. The $N$ = 3 $\to$ 2 transition is brightest close to Her 36 at ($\Delta\alpha$ = 0$\arcsec$, $\Delta\delta$ = 12$\arcsec$) with the emission extending toward north-west of Her 36. The hot gas near Her 36 would be able to excite C$_{2}$H molecule to rotational levels with higher upper level energies whereas the gas deep into the molecular cloud would be comparatively cooler. It is noteworthy that the extended emission, as shown in Fig.~\ref{fig:ci-co-c2h}, toward the north-west of Her 36 is also seen in low and mid-$J$ CO transitions and in \ci\ observed toward M8 by the APEX and IRAM~30~m telescopes \citep{refId0}.\ 

In order to investigate the relation of the dense and cold molecular cloud material to the hot ionized gas in M8, we compared our observed C$_{2}$H data with observations obtained from 2 surveys conducted at different wavelengths. Firstly, we used the 8~$\mu$m data from the Galactic Legacy Infrared Mid-Plane Survey Extraordinaire (GLIMPSE, \citealt{2003PASP..115..953B}) archive observed with Spitzer Space Telescope. The 8~$\mu$m band of the Infrared Array Camera (IRAC) used by Spitzer covers emission from polycyclic aromatic hydrocarbons (PAHs) which are small molecules excited by strong UV radiation. Since the IR emission from these large molecules (or small dust grains) constitutes fluorescence that results from far-UV pumping, it emerges from the surface layers of dense molecular clouds, PAH emission probes recent high-mass star formation \citep{Tielens2008}. Figure~3 shows the velocity integrated intensity maps of C$_{2}$H 1 $\to$ 0 and 3 $\to$ 2 transitions in color scale overlaid with GLIMPSE 8~$\mu$m contours in black, which peaks at Her 36 ($\Delta\alpha$ = 0$\arcsec$, $\Delta\delta$ = 0$\arcsec$) and at the \cii\ emission peak ($\Delta\alpha$ = 30$\arcsec$, $\Delta\delta$ = --2$\arcsec$). \cii\ was observed in May 2016 by the SOFIA telescope and the data has been reported in \citep{refId0}. Secondly, we obtained, from the Herschel Space Archive\footnote{https://irsa.ipac.caltech.edu/applications/Herschel/} (HSA), data collected with the Spectral and Photometric Imaging Receiver (SPIRE, \citealt{2010A&A...518L...3G}) aboard the Herschel Space Observatory \citep{2010A&A...518L...1P}. The SPIRE data trace interstellar dust. The white contours in Fig.~3 represent the SPIRE 250~$\mu$m continuum emission, which peaks at Her 36 ($\Delta\alpha$ = 0$\arcsec$, $\Delta\delta$ = 0$\arcsec$) and is extended in north-west direction with a secondary peak coinciding the peak of the C$_{2}$H 1 $\to$ 0 emission ($\Delta\alpha$ = --56$\arcsec$, $\Delta\delta$ = 28$\arcsec$).                

\begin{figure}[htp]
 \centering
 \subfigure{\includegraphics[width=80mm]{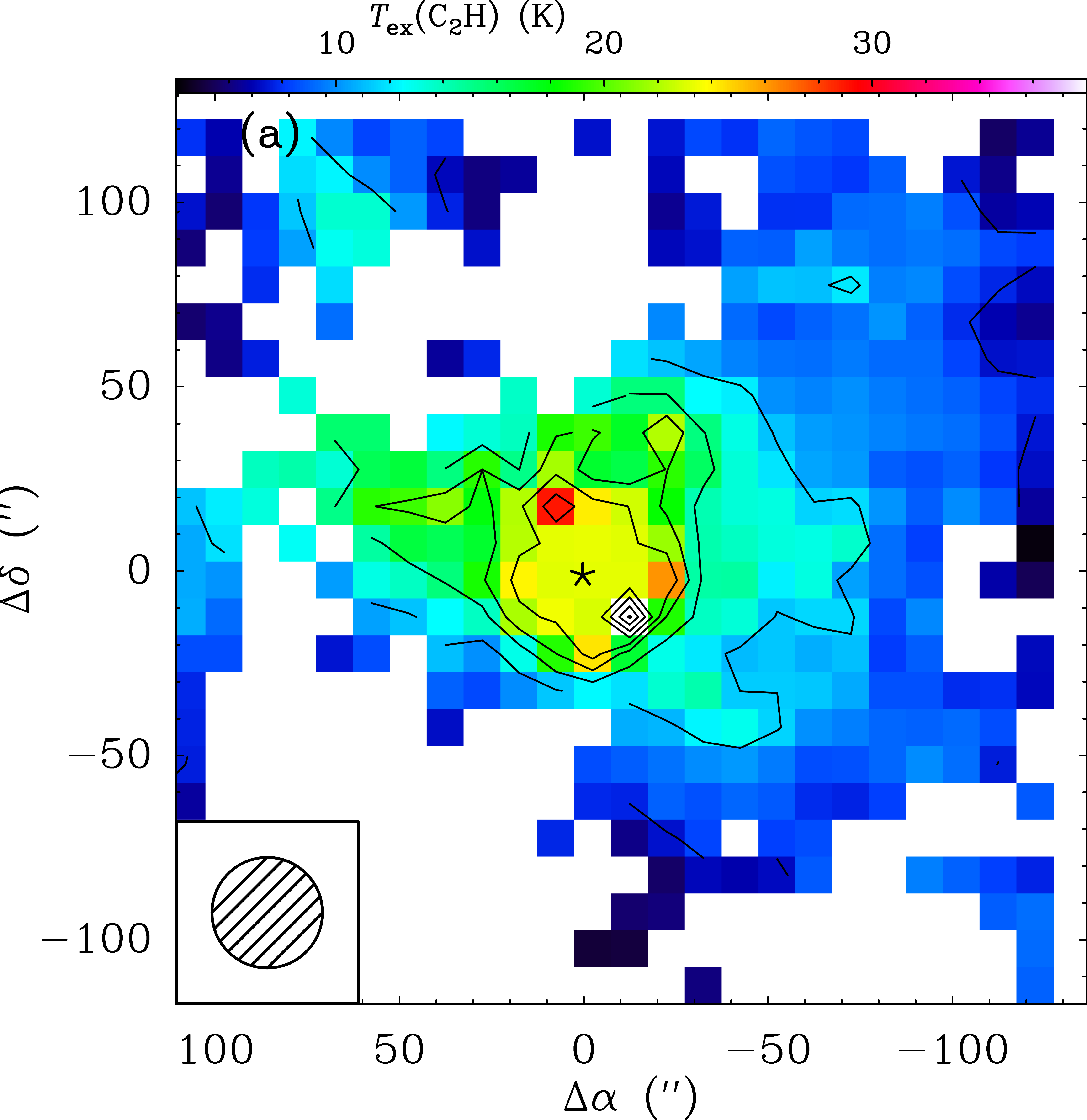}}\quad
 \subfigure{\includegraphics[width=80mm]{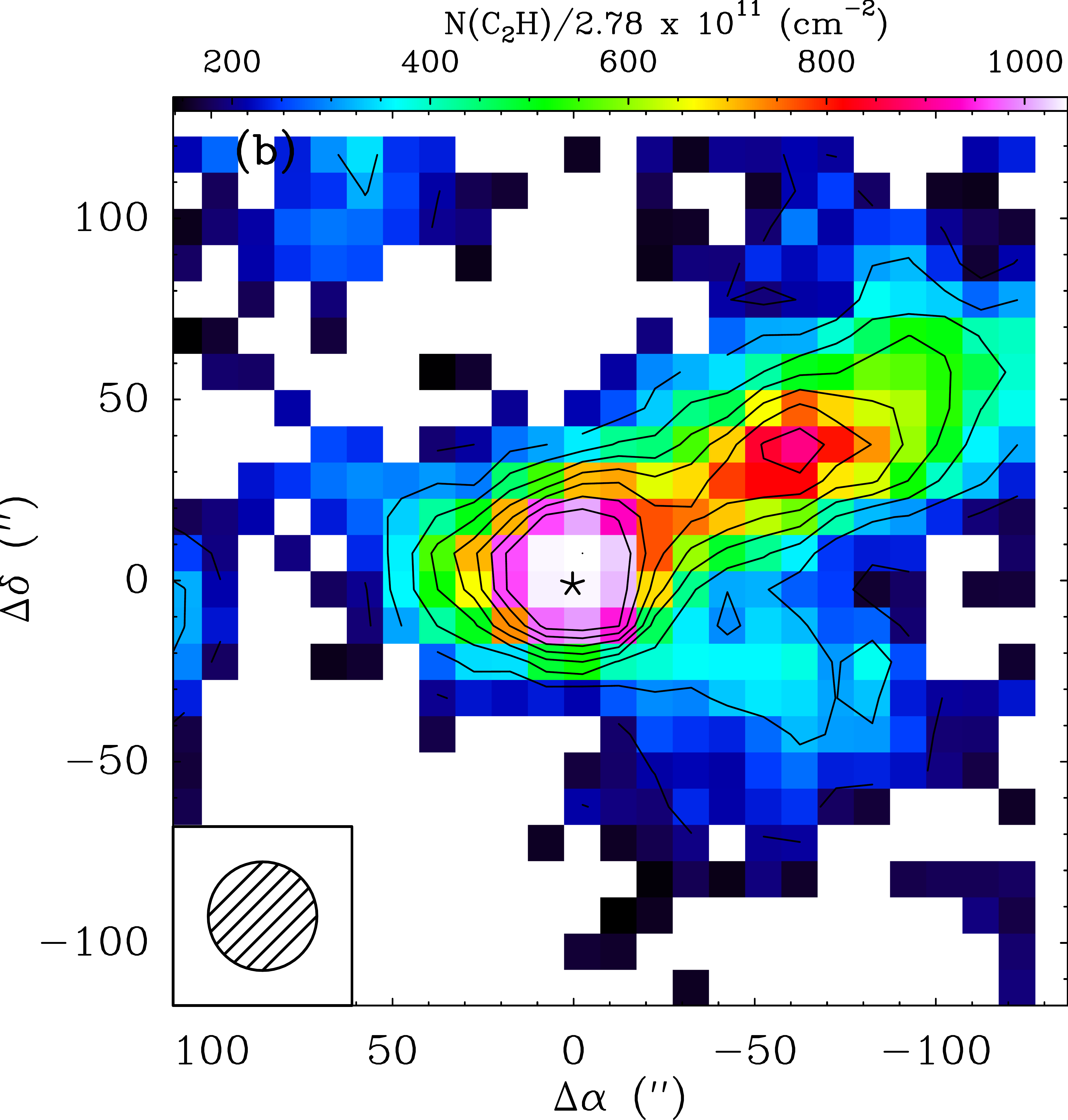}}

  \caption{(a, upper panel) shows the excitation temperature defining the populations of $N$ = 1 and 3 levels and (b, lower panel) shows the total column density of C$_{2}$H. The asterisk represents Her 36 which is the central position ($\Delta\alpha$ = 0, $\Delta\delta$ = 0) at R.A.(J2000) = 18\textsuperscript{h}03\textsuperscript{m}40.3\textsuperscript{s} and Dec.(J2000) = --24$\degree$22$\arcmin$43$\arcsec$. The contour levels are 10\% to 100\% in steps of 10\% of the corresponding peak emissions. The values of main beam brightness temperatures and that of the velocity integrated intensities for the $N$ = 1 and 3 levels used to calculate the $T_{\rm ex}$(C$_{2}$H) and $N$(C$_{2}$H) were extracted from maps convolved to the same resolution of 30$\arcsec$.}
   \label{tex_colden} 
\end{figure}

\section{Analysis}

\begin{figure*}[htp]
 \centering
 \subfigure{\includegraphics[width=58mm]{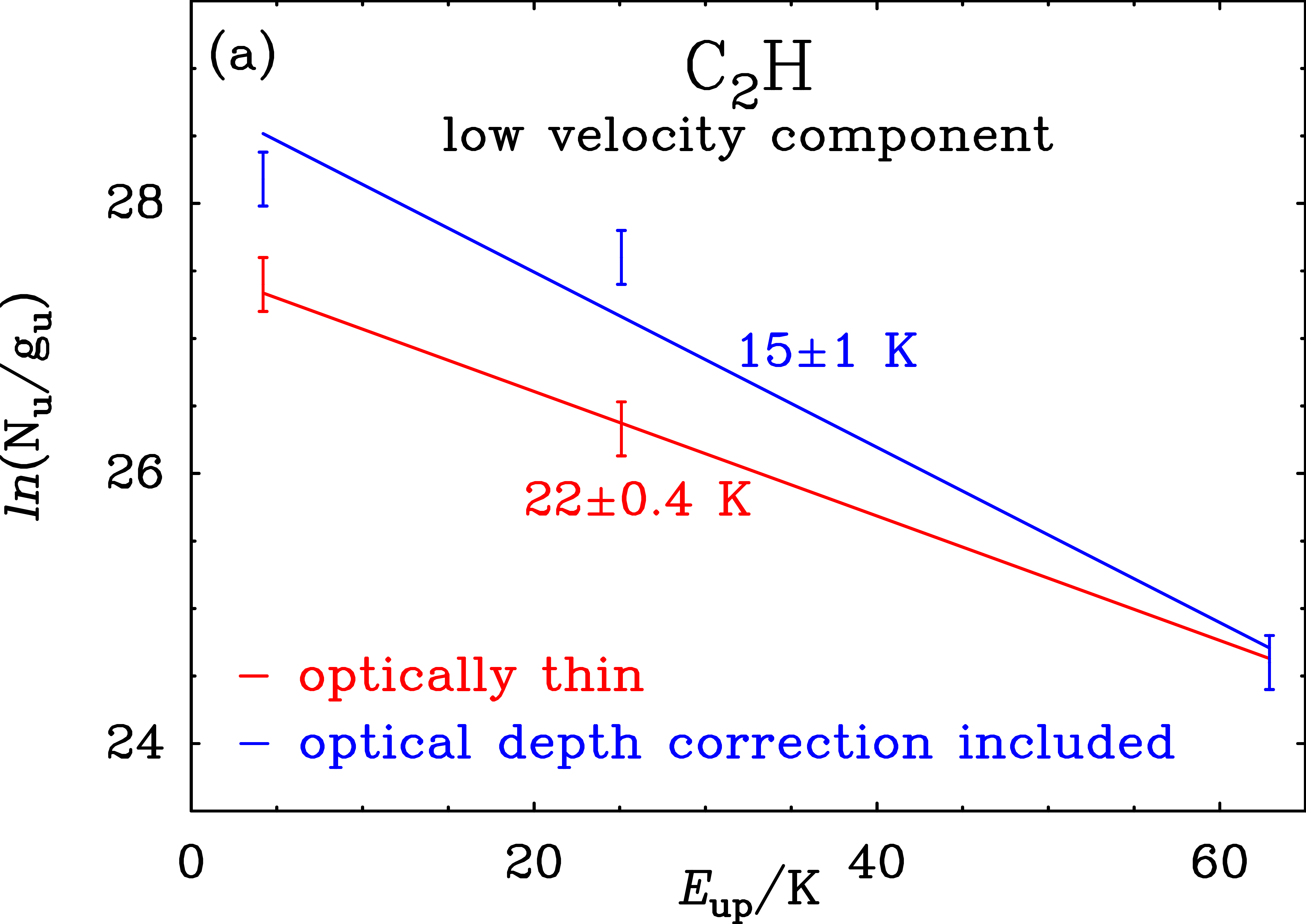}}\quad
 \subfigure{\includegraphics[width=58mm]{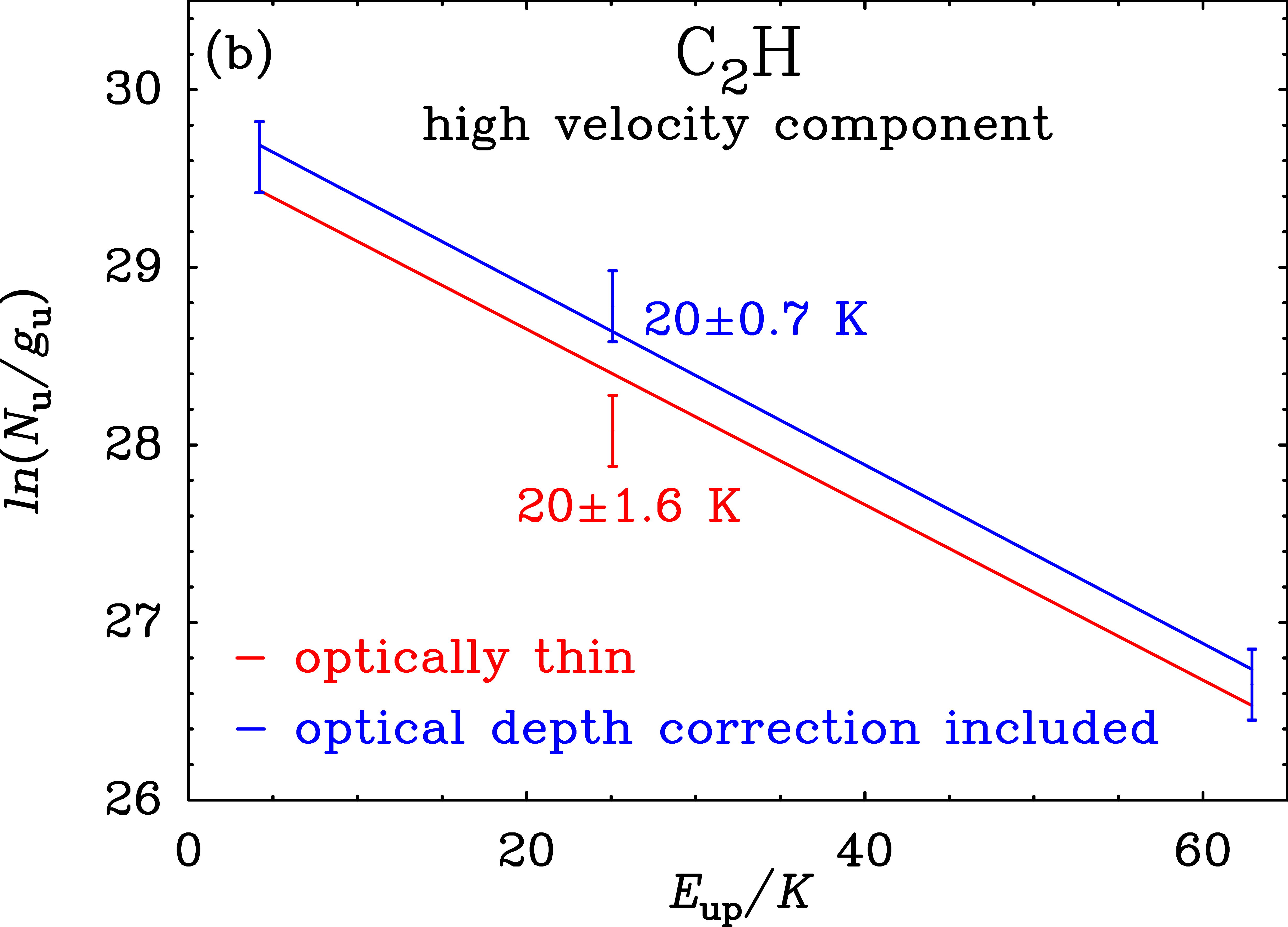}} 
 \subfigure{\includegraphics[width=58mm]{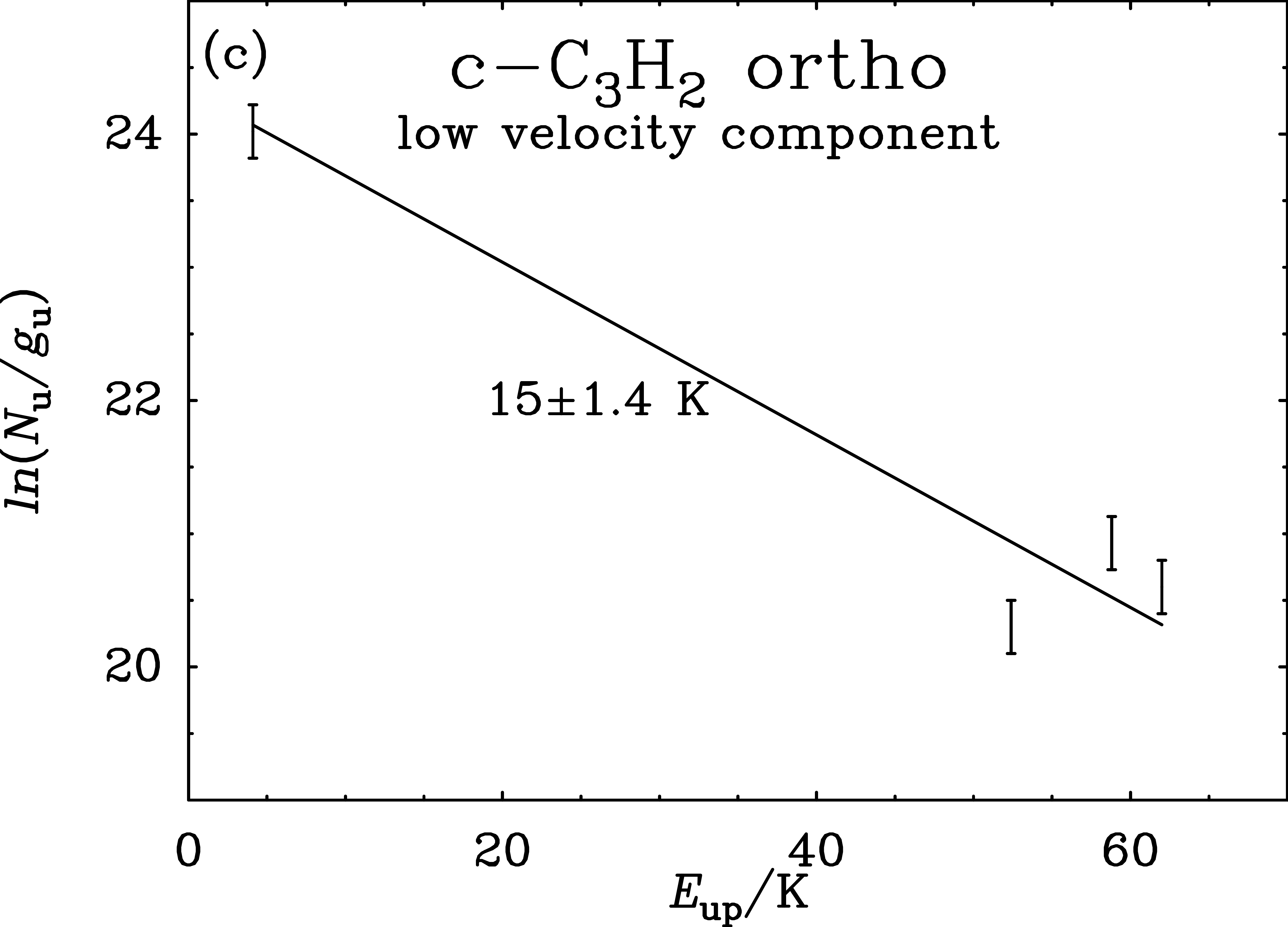}}
 \subfigure{\includegraphics[width=58mm]{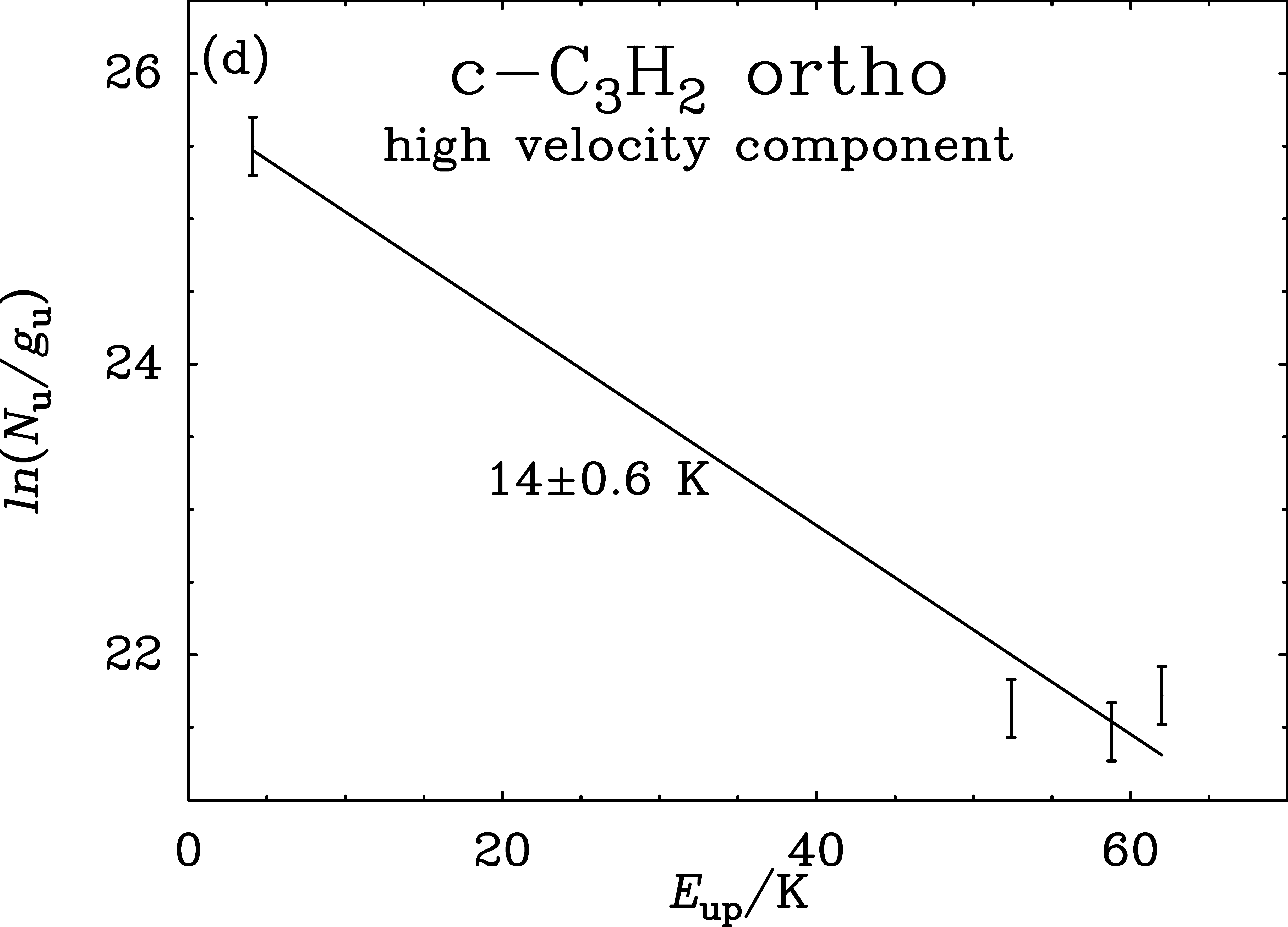}}\quad
 \subfigure{\includegraphics[width=58mm]{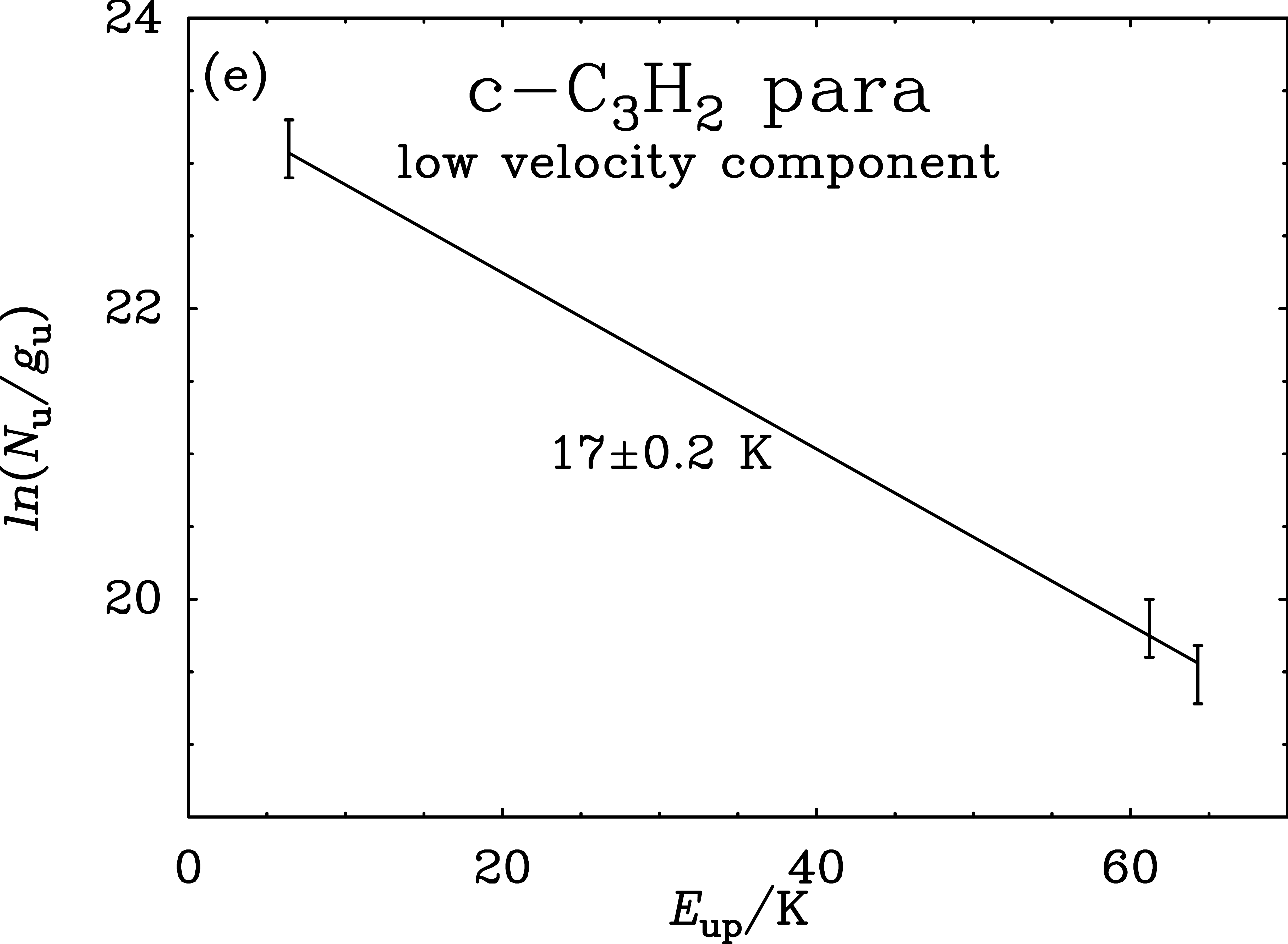}}\quad
 \subfigure{\includegraphics[width=58mm]{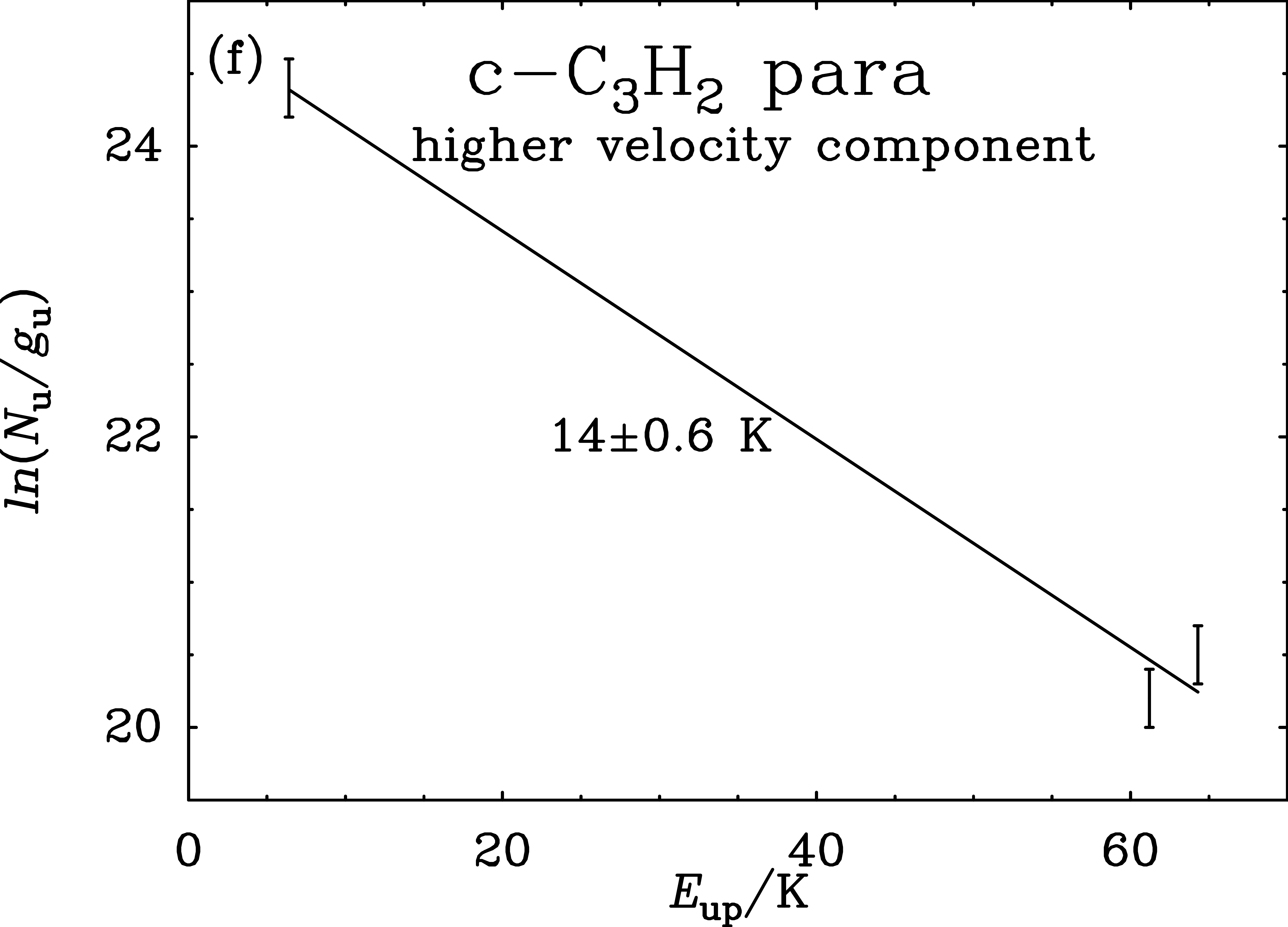}}

\caption{Rotational diagrams of various transitions of the $N$ = 1 $\to$ 0, 3 $\to$ 2 and 5 $\to$ 4 transitions of C$_{2}$H (top row); $J$ = 2 $\to$ 1, 6 $\to$ 5, 7 $\to$ 6 and 8 $\to$ 7 transitions of ortho c-C$_3$H$_2$ (top and bottom rows); and $J$ = 2 $\to$ 1, 7 $\to$ 6 and 8 $\to$ 7 transitions of para c-C$_3$H$_2$ (bottom row) observed toward Her 36. For C$_{2}$H, rotational diagrams are plotted with an assumption of optically thin C$_{2}$H (in red) and with including the optical depth correction factors (in blue). Fitted values of rotational temperatures are given for each molecule. The error bars were calculated from the maximum noise of the integrated intensities of individual transitions and from calibration uncertainties of 20\%.}
 \label{rot_dia} 
\end{figure*}

Various transitions of C$_{2}$H and c-C$_{3}$H$_2$ observed with the IRAM 30~m and APEX~12~m telescopes allow us to carry out a detailed analysis to determine the temperature and density of the gas responsible for the emission of hydrocarbons in M8 with several complementary methods. We started with estimating the excitation temperature and the total C$_{2}$H column density by using its OTF maps of 1 $\to$ 0 and 3 $\to$ 2 transitions in Sect.~4.1. We then calculated column densities and abundances of both C$_{2}$H and c-C$_{3}$H$_2$ from rotational diagrams in Sect.~4.2. To complete the analysis, we made use of a radiative transfer model to constrain the H$_2$ densities in M8 in Sect.~4.3 and compared our observed results with PDR models in Sect.~4.4.     

\begin{table*}[ht]
\small
\caption{Physical parameters calculated from rotational diagrams.}
\centering
\begin{threeparttable}
\begin{tabular}{c c c c c c}
\hline\hline
\noalign{\smallskip}
\multicolumn{1}{c}{Species} & \multicolumn{2}{c}{Lower velocity component} & \multicolumn{2}{c}{Higher velocity component} & \multicolumn{1}{c}{Abundance}\\
\hline
\noalign{\smallskip}
 & $T_{\rm rot}$ (K) & $N(\rm X)$ (cm$^{-2}$) & $T_{\rm rot}$ (K) & $N(\rm X)$ (cm$^{-2}$) & $N(\rm X)$/$N(\rm H_{2})$\\
\hline
\noalign{\smallskip}
C$_{2}$H(optically thin) & 22 $\pm$ 0.4 & (3.8 $\pm$ 0.2) $\times$ 10$^{13}$ & 20 $\pm$ 1.6 & (2.9 $\pm$ 0.1) $\times$ 10$^{14}$ & (0.9 $\pm$ 0.04) $\times$ 10$^{-8}$ \\
C$_{2}$H(optical depth correction included) & 15 $\pm$ 1 & (9.8 $\pm$ 0.5) $\times$ 10$^{13}$ & 20 $\pm$ 0.7 & (3.8 $\pm$ 0.3) $\times$ 10$^{14}$ & (1.3 $\pm$ 0.1) $\times$ 10$^{-8}$\\

c-C$_{3}$H$_{2}$ - ortho & 15 $\pm$ 1.4 & (4 $\pm$ 0.1) $\times$ 10$^{12}$ & 14 $\pm$ 0.6 & (1.8 $\pm$ 0.43) $\times$ 10$^{13}$ & (6 $\pm$ 1.2) $\times$ 10$^{-10}$ \\
 c-C$_{3}$H$_{2}$ - para & 17 $\pm$ 0.2 & (2.2 $\pm$ 0.1) $\times$ 10$^{12}$ & 14 $\pm$ 0.6 & (7.2 $\pm$ 0.2) $\times$ 10$^{12}$ & (2.5 $\pm$ 0.1) $\times$ 10$^{-10}$ \\
\hline
\noalign{\smallskip}

\end{tabular}

\end{threeparttable}
\label{rot_dia}
\end{table*}

\subsection{Excitation temperature and column density estimates of C$_{2}$H} 
Assuming a beam filling factor of unity and optically thin emission in both the C$_{2}$H $N$ = 1 $\to$ 0, $J$ = 3/2 $\to$ 1/2, $F$ = 2 $\to$ 1 at 87.316~GHz and $N$ = 3 $\to$ 2 $J$ = 7/2 $\to$ 5/2, $F$ = 4 $\to$ 3 at 262.004~GHz transitions, the excitation temperature can be estimated by:

\begin{equation}
   T_{\rm ex} = \frac{-20.95}{ln\big(0.144\frac{W_{\rm 3}}{W_{\rm 1}})} \rm K\,, 
\end{equation}\\
where $W_{\rm 1}$ and $W_{\rm 3}$ are the velocity integrated intensities in K~cm~s$^{-1}$ of $N$ = 1 $\to$ 0 and 3 $\to$ 2 transitions of C$_{2}$H respectively. The resulting $T_{\rm ex}$ distribution is shown in Fig.~\ref{tex_colden}~(a), and these are only lower limits to the excitation temperature as the beam filling factor is assumed to be unity. The peak of the $T_{\rm ex}$ distribution $\sim$ 38~K~$\pm$ 11~K lies in the immediate south-west of Her 36 ($\Delta\alpha$ = -15$\arcsec$, $\Delta\delta$ = -13$\arcsec$delta) and $T_{\rm ex}$ decreases with the distance from the star.\

Assuming the environment to be in Local Thermodynamical Equilibrium (LTE) such that all transitions have the same excitation temperature and that $T_{\rm ex}$ equals the molecule's rotation temperature, $T_{\rm rot}$, we calculated the total column density of C$_{2}$H using the computed $T_{\rm ex}$ and the velocity integrated intensity, $W_{\rm 3}$, of $N$ = 3 $\to$ 2 transition of C$_{2}$H:

\begin{equation}
     N(\rm C_{2}H) = 2.78 \times 10^{11} (1.9~\textit{T}_{\rm ex} + 1.38)~exp\Bigg(\frac{25.15}{\textit{T}_{\rm ex}}\Bigg)~\textit{W}_{\rm 3}~cm^{-2}\,.
 \end{equation}\\

Figure~\ref{tex_colden}~(b) shows the resulting C$_{2}$H total column density distribution with a peak value of $N(\rm C_2H)$ $\sim$ (3 $\pm$ 0.26) $\times$ 10$^{14}$ cm$^{-2}$ at Her 36 and a secondary peak toward the north-west of Her 36 in the molecular cloud similar to the secondary ATLASGAL peak. This results in an abundance of $N(\rm C_{2}H)$/$N(\rm H_{2})$ $\sim$ 8 $\times$ 10$^{-8}$, where we adopted $N(\rm H_{2})$ $\sim$ 3.75 $\times$ 10$^{22}$ cm$^{-2}$ \citep{refId0}.  

\subsection{Rotational Diagrams of C$_{2}$H and c-C$_{\rm 3}$H$_{\rm 2}$}
In Sect.~4.1, we presented the maps of excitation temperature and total column density distribution of C$_2$H in M8, around Her 36. But c-C$_3$H$_2$ and the higher transition of C$_2$H were a part of point observations done by integrating deeply toward Her 36. So, we exploit the technique of rotational diagrams to perform a detailed analysis of more transitions as compared to Sect.~4.1. Also, here we plot rotational diagrams for both low and high velocity components of all observed transitions of C$_2$H and c-C$_3$H$_2$ observed toward Her 36. Assuming LTE, rotational diagrams (\enquote{Boltzmann plots}) can be used to estimate the total column densities, $N_{\rm tot}$, of C$_2$H and c-C$_3$H$_2$ and the rotation temperatures, $T_{\rm rot}$ describing their level populations \citep{1999ApJ...517..209G}. In a rotational diagram, $ln(N_{\rm u}/g_{\rm u}$) is plotted vs. $E_{\rm up}/k$. Here, $ln(N_{\rm u}$) and $g_{\rm u}$ ($\equiv 2J+1$) are the column density in and the degeneracy of the upper energy level, respectively, $E_{\rm up}$ is the upper level energy and $k$ is the Boltzmann constant.\


\begin{figure*}[htp]
  \centering
 \subfigure{\includegraphics[width=80mm]{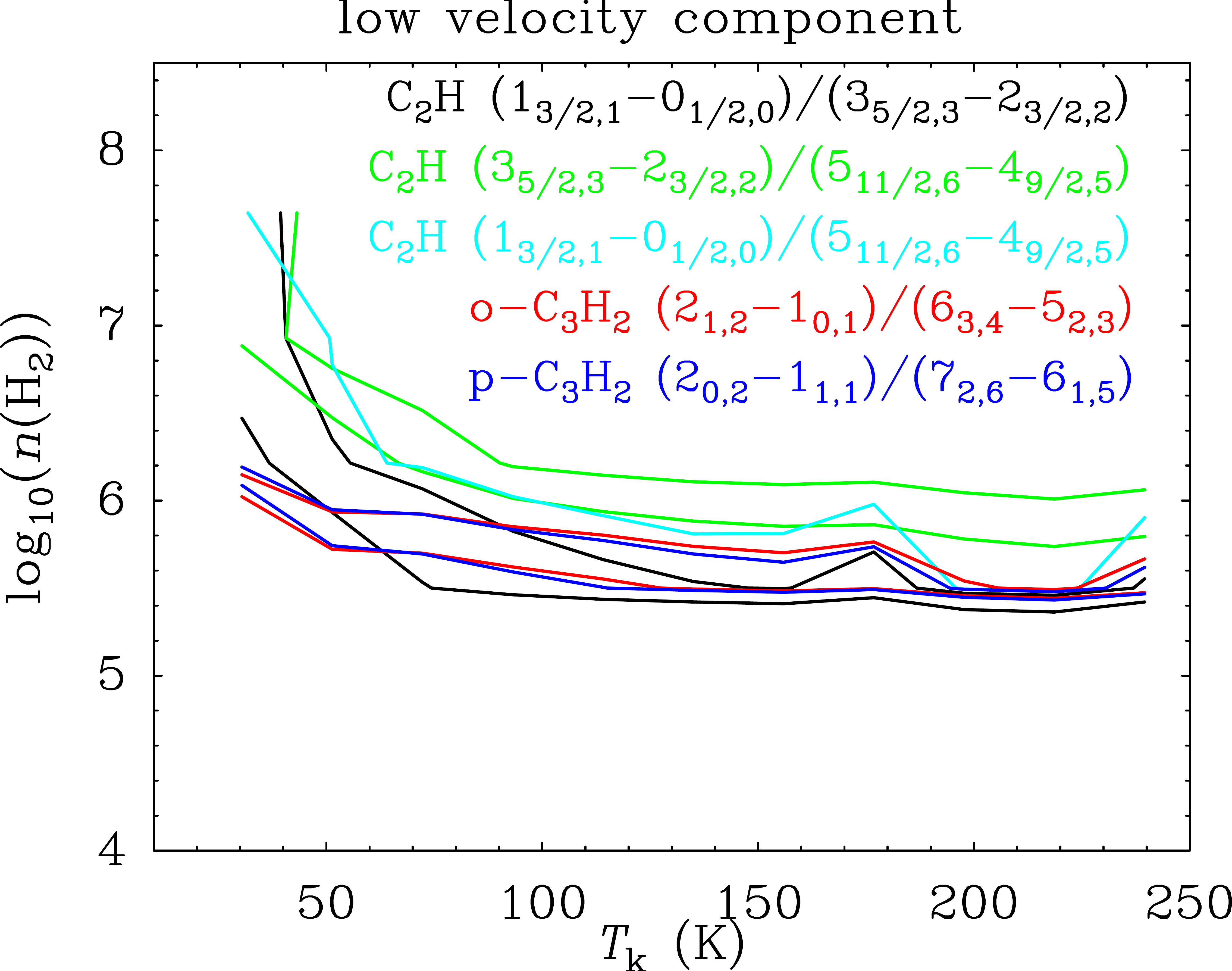}}\quad
 \subfigure{\includegraphics[width=80mm]{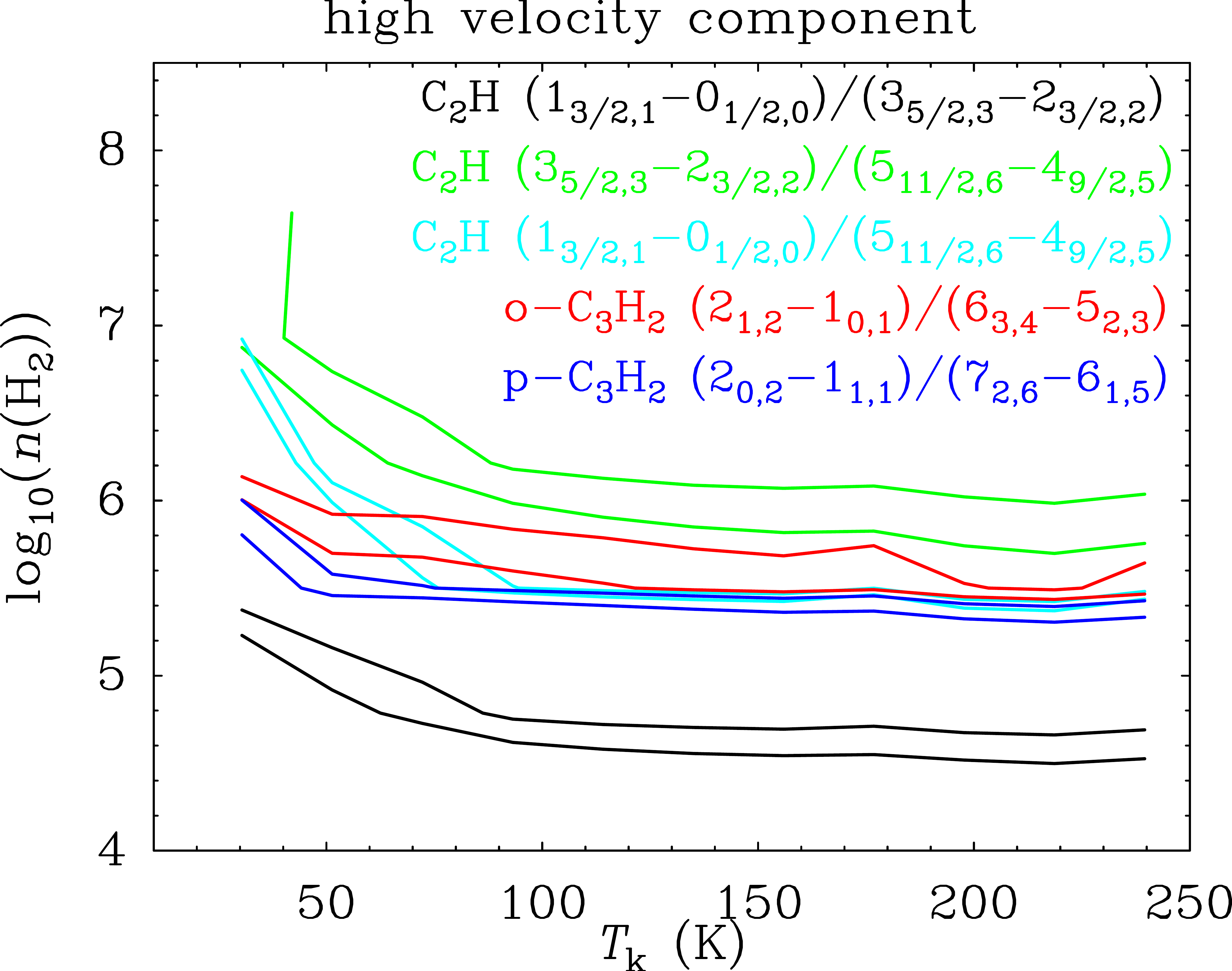}}

  \caption{Results obtained from RADEX modeling for the main beam brightness temperature ratios of various transitions of C$_{2}$H and c-C$_{3}$H$_{2}$ toward Her 36 in a log$_{10}$(n(H$_{2}$)) vs $T_{\rm k}$ grid. The left and right panels correspond to the low velocity and the high velocity components respectively. The two contours per color represent the RADEX modeling output values equal to the upper and lower bounds of the value obtained for each ratio from the observations with an error of 20\%.}
  
\end{figure*}

Given the extended emission seen in the velocity integrated intensity maps in Fig.~\ref{vel_int_maps}, we adopted a beam filling factor of 1 to make rotational diagrams. Moreover, we assume the hfs lines of each rotational transition to be treated as a single component with a quantum number $N$. We then calculated the total velocity integrated intensity, the upper level degeneracy $g_{\rm u}$ and the line strength of different transitions as the sum of all the observed hfs components of each $N$ + 1 $\to$ $N$ transition. Using the weighted average with the relative strength of each line as weight, the hfs-weighted frequency $\nu$ was obtained (as in \citealt{2015A&A...575A..82C}) and the Einstein coefficient $A$ was calculated using \citet[Eq.~25]{1999ApJ...517..209G}.\ 

\textbf{Figures~5~(a) and~(b)} show rotational diagrams of C$_{2}$H for low (2--8~km~s$^{-1}$) and high (8--15~km~s$^{-1}$) velocity components respectively. Firstly, under the assumption that the C$_{2}$H lines are optically thin, i.e. the optical depth correction term $C_{\rm \tau}$ is unity in \citet[Eq.~24]{1999ApJ...517..209G}, we plotted $N^{\rm thin}_{\rm u}$/$g_{\rm u}$ vs. $E_{\rm up}$/$k$ in red for three transitions toward Her 36. Since the low velocity component could only be spectrally resolved for the $N$ = 5 $\to$ 4 transition, we estimated the low velocity component's line parameters for the other transitions assuming the same velocity and line width as that of $N$ = 5 $\to$ 4. Subsequently, we plotted the corrected diagrams, in blue, by including the optical depth correction values according to \citet[Eq.~24]{1999ApJ...517..209G}. The optical depth values were obtained from the hfs fit and are listed in Table~\ref{line_par_cch}.\

Figures~5~(c),~(d),~(e) and~(f) show rotational diagrams of the ortho and para species of c-C$_{3}$H$_{2}$ for the low (2--8~km~s$^{-1}$) and high (8--15~km~s$^{-1}$) velocity components. We first assumed c-C$_{3}$H$_{2}$ to be optically thin and from $J_{\rm K_{\rm a}, K_{\rm b}}$ = 2$_{1,2}$ $\to$ 1$_{0,1}$ ortho and $J_{\rm K_{\rm a}, K_{\rm b}}$ = 2$_{0,2}$ $\to$ 1$_{1,1}$ para transitions, we calculated the ortho/para ratio $\sim$ 2.2 for low velocity component and 3.3 for high velocity component. These ratios were used to estimate the contributions from ortho and para species to the $J_{\rm K_{\rm a}, K_{\rm b}}$ = 7$_{1,6}$ $\to$ 6$_{2,5}$ and 8$_{1,8}$ $\to$ 7$_{0,7}$ ortho; and 7$_{2,6}$ $\to$ 6$_{1,5}$ and 8$_{0,8}$ $\to$ 7$_{1,7}$ para transitions, where the lines from the two species are blended together as can be seen in Fig.~2. From the obtained rotational temperatures, we estimated the column densities of C$_{2}$H and c-C$_{3}$H$_{2}$, which in turn were used to determine the optical depths of c-C$_{3}$H$_{2}$ \citep[Eq.~27]{1999ApJ...517..209G}. The estimated optical depth values came out to be very low, validating our assumption of optically thin c-C$_{3}$H$_{2}$.\        

The calculated rotational temperatures $T_{\rm rot}$ and column densities $N(\rm X)$ along with their abundances $N(\rm X)$/$N(\rm H_{2})$ of C$_{2}$H and c-C$_{3}$H$_{2}$ are given in Table~\ref{rot_dia}. The total column density value of C$_2$H at the Her 36 position calculated using the rotational diagrams is very similar to that determined in Sect.~4.1.  

\subsection{Non-LTE calculations}
In Sects.~4.1 and 4.2, we obtained the column densities and abundances of C$_{2}$H, o-C$_{3}$H$_{2}$ and p-C$_{3}$H$_{2}$ under the assumption of LTE. The derived rotational temperatures ($\sim 15$--$20$~K) are much lower than values that one might expect in the vicinity of Her 36 and temperatures derived in our previous study \citep{refId0}. This could be caused by the fact that the critical densities of the higher $E_{\rm up}$ lines of C$_2$H and C$_3$H$_2$ are much higher than those of the lower $E_{\rm up}$ lines of both molecules. This leads to subthermal excitation of the higher $E_{\rm up}$ lines and, consequently, to lower than LTE integrated intensities (and lower $N_{\rm up}$/$g_{\rm up}$ values). This results in a steepening of the fitted line whose slope gives $T_{\rm rot}$ and an underestimation of that quantity. Therefore the resultant rotation temperature and column densities should only be considered as guiding values. To address this issue, we employ a non-LTE analysis that delivers kinetic temperature and H$_2$ volume densities of the gas that gives rise to the emission of C$_{2}$H and C$_{3}$H$_{2}$.
RADEX is a non-LTE radiative transfer program \citep{2007A&A...468..627V}, which uses the escape probability approximation for a homogeneous medium and takes into account optical depth effects. We chose a uniform sphere geometry. The collision rates used in the modeling are provided by the Leiden Molecular and Atomic Database\footnote{http://www.strw.leidenuniv.nl/moldata/} (LAMDA; \citet{2005A&A...432..369S}). The C$_{2}$H-H$_{2}$ collision rates are calculated by multiplying a factor of 1.36 to the C$_{2}$H-He collision rates given by \citet{2012MNRAS.421.1891S}, while the C$_{3}$H$_{2}$-H$_2$ collision rates are given by \citet{2000A&AS..142..113C}. We computed grids in kinetic temperatures in the range 20--250~K and H$_{2}$ volume densities in the range 10$^{3}$--10$^{8}$~cm$^{-3}$ and a background temperature of 2.73~K as input parameters and then calculated brightness temperature ratios of different transitions of C$_{2}$H, o-C$_{3}$H$_{2}$ and p-C$_{3}$H$_{2}$.\

\begin{figure*}[htp]
  \centering
 \subfigure{\includegraphics[width=90mm]{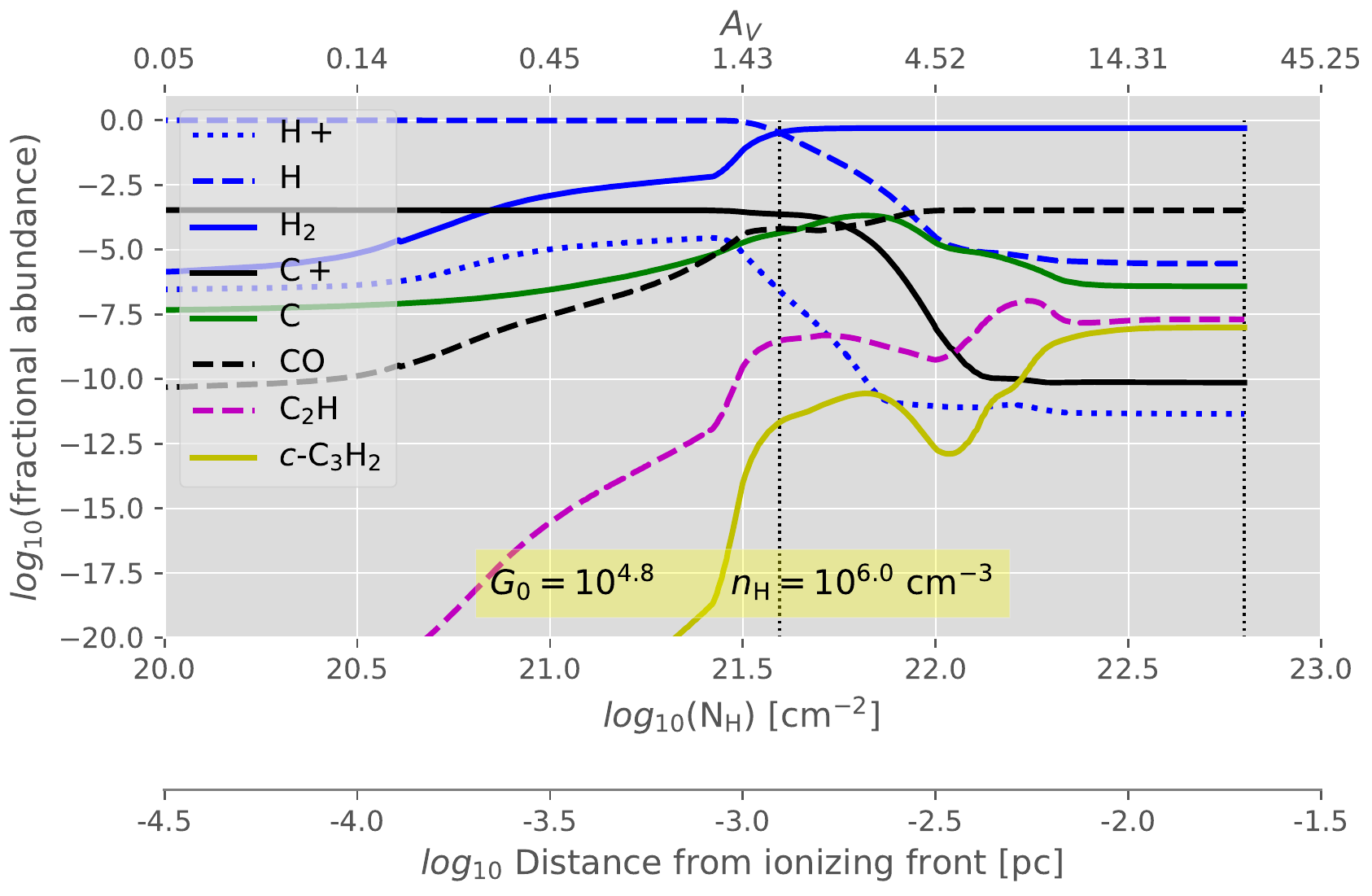}}\quad
 \subfigure{\includegraphics[width=90mm]{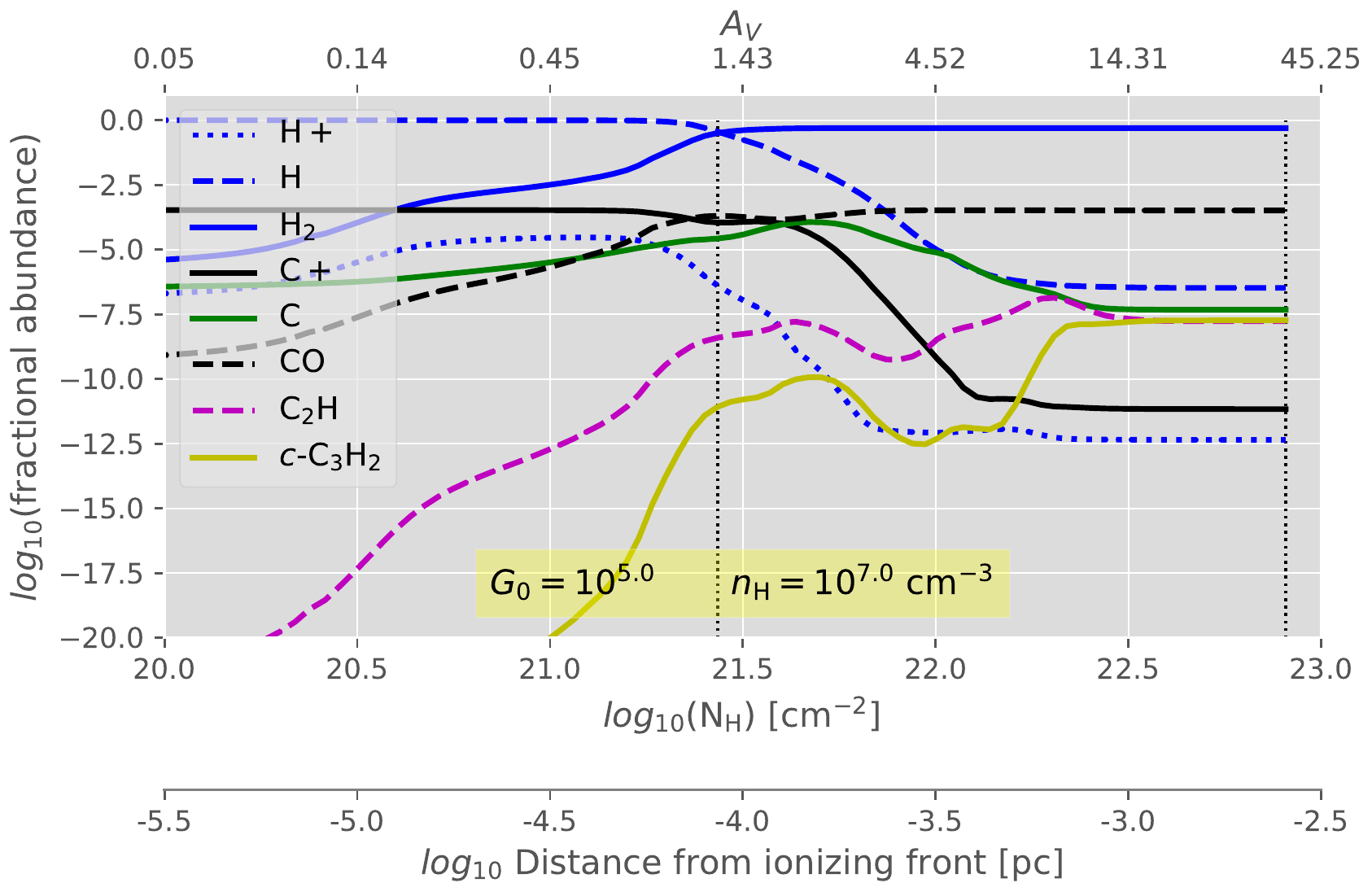}}

  \caption{PDR models for M8 for $G_0$ = $10^{4.8}$ and $n_{\rm H}$ = 10$^{5}$~cm$^{-3}$ (left panel) and $G_0$ = $10^{5}$ and $n_{\rm H}$ = 10$^{7}$~cm$^{-3}$ (right panel). The fractional abundances ($n_X / n_{\rm H}$) are represented by the color codes/lines patterns indicated on the plots and measured by the ordinates. The top abscissa indicates the corresponding visual extinction ($A_{\rm V}$) as seen from the ionizing front (the impinging radiation field goes from left to right of the plot). The first bottom abscissa gives the corresponding total hydrogen column density $N(\rm H)$. A second bottom abscissa shows the actual distance (in pc) from the ionizing front. The vertical dashed lines demarcate the region where most of the hydrogen gas is in molecular form.}
  \label{fig:PDR-models}
\end{figure*}

We adopted the line widths from the average spectra of our data of C$_{2}$H, o-C$_{3}$H$_{2}$ and p-C$_{3}$H$_{2}$ i.e. 3, 2.5, 2.2~km~s$^{-1}$ respectively for the low velocity (2--8 km~s$^{-1}$) component and 3.6, 3.3, 3~km~s$^{-1}$ respectively for the high velocity (8--15~km~s$^{-1}$) component. For modeling, we used the C$_{2}$H, o-C$_{3}$H$_{2}$ and p-C$_{3}$H$_{2}$ column densities as estimated from the rotational diagrams in Sect.~4.2 i.e. 7 $\times$ 10$^{13}$, 7 $\times$ 10$^{12}$ and 4 $\times$ 10$^{12}$~cm$^{-2}$ respectively for the low velocity component and 4 $\times$ 10$^{14}$, 2 $\times$ 10$^{13}$ and 7 $\times$ 10$^{12}$~cm$^{-2}$ respectively for the high velocity component. Figure~6 shows the brightness temperature ratios (in different colors) of various transitions of C$_{2}$H, o-C$_{3}$H$_{2}$ and p-C$_{3}$H$_{2}$ on a log$_{10}$($n$(H$_{2}$)) vs. $T_{\rm kin}$ plot. The two different contours per color represent the RADEX modeling output values equal to the upper and lower bounds of the value obtained for each ratio from the observations with an error of 20$\%$. For low $T_{\rm kin}$ (20--100~K), a gradient in density is seen, which then almost saturates for higher $T_{\rm kin}$ (100--250~K).\

For very low temperatures $T_{\rm kin}$ $<$ 50~K, the H$_{2}$ volume density $n(\rm H_{2})$ is as high up to 5 $\times$ 10$^{7}$~cm$^{-3}$. But in our previous work, we determined kinetic temperatures $T_{\rm kin}$ $\sim$ 100--150~K in M8 \citep[Sect.~4.3]{refId0}, which puts a constraint on the volume density $n(\rm H_{2}$) values. For the low velocity component shown in Fig.~6~(a), all hydrocarbons probe similar gas that is a part of the foreground veil, which has been found by and discussed in \citet[Sect.~5.1][Fig.~15]{refId0}. The foreground veil accelerating toward us by the strong radiation and wind of Her 36, which is a part of the cold dense molecular cloud has H$_{2}$ volume density $n(\rm H_{2})$ in a range of 5 $\times$ 10$^{5}$--5 $\times$ 10$^{6}$~cm$^{-3}$. For the high velocity component shown in Fig.~6~(b), the different ratios of C$_{2}$H probe gas of different densities. For $T_{\rm kin}$ $\sim$ 100--150~K, C$_{2}$H (1~$\to$~0)/(5 $\to$ 4) probe the same gas as o-C$_{3}$H$_{2}$ and p-C$_{3}$H$_{2}$ with $n(\rm H_{2})$ $\sim$ 1 $\times$ 10$^{6}$~cm$^{-3}$, while C$_{2}$H (3~$\to$~2)/(5 $\to$ 4) probe slightly denser gas with $n(\rm H_{2})$ $\sim$ 5 $\times$ 10$^{6}$~cm$^{-1}$ and C$_{2}$H (1~$\to$~0)/(3~$\to$~2) probe lower density gas with $n(\rm H_{2})$ $\sim$ 5 $\times$ 10$^{4}$~cm$^{-3}$. This velocity component with an H$_{2}$ volume density $n(\rm H_{2})$ in the range of 5 $\times$ 10$^{4}$--5 $\times$ 10$^{6}$~cm$^{-3}$, consists of the gas very close to Her 36 and also the background material toward north-east of it \citep[Sect.~5.1]{refId0}.

\subsection{Comparison with PDR models}
In order to compare the results obtained with the LTE and non-LTE models, we use PDR models to estimate the relative abundances of the C$_{2}$H and C$_3$H$_2$ species. These PDR models are based on \citet{2005A&A...436..397M} which include more than 300 species in the chemical network, photo-electric heating from PAHs and small dust grains, cosmic-ray heating, and other classical heating and cooling mechanism for the thermal and chemical balance. A semi infinite slab geometry and irradiation from one side without geometrical dilution are assumed. Note that the PDR models do not distinguish between ortho and para molecules, but they provide the total abundance of C$_3$H$_2$ (and all other available species) self-consistently depending on the depth and visual extinction of the parallel slab. The model uses total gas density $n_{\rm H}$ = $n(\rm H)$ + 2$n(\rm H_2)$, as an input parameter and the actual density of the collisional partner, molecular hydrogen, $n(\rm H_2)$ is calculated self-consistently along with the densities of other species from chemical and thermal balance equations, and is depth dependent.\  

We model four scenarios with two different total gas densities, $n_{\rm H}$, and two levels of strength of the impinging radiation field, $G_0$, in Habing units. In the models we assume solar metallicity and the visual extinction is estimated using the up-to-date relation $N({\rm H})=2.21\times 10^{21}~{\rm cm}^{-2}\times A_{\rm V}$ found by \citet{2009MNRAS.400.2050G}.

For $G_0$, we used the values $0.6\times 10^5$ and $1.12 \times 10^5$ estimated in our previous paper \citep[Sect.~5.2]{refId0}, using \citet[Eq.~9.2]{Tielens2008} for an O star with electron density, $n_{\rm e}$, of 2000--4000~cm$^{-3}$ and electron temperature, $T_{\rm e}$, of 7000--9000~K \citep{1986AJ.....91..870W, 1999ApJS..120..113E}. We need to use quite high total gas densities $n_{\rm H}$ of $10^6$ and $10^7$~cm$^{-3}$ in order to reach densities of the collision partner $n(\rm H_2)$ similar to the values found in Sect. 4.3 using RADEX. Figure~\ref{fig:PDR-models} shows the models with the lowest density and impinging radiation field (\textit{left}) and with the highest density and radiation field (\textit{right}) tested.\ 

For the highest density model ($n_{\rm H} = 10^7$~cm$^{-3}$) we get relative abundances between $4\times 10^{-9}$ and $2\times 10^{-8}$ for C$_{2}$H and between $8\times 10^{-8}$ and $2\times 10^{-8}$ for C$_3$H$_2$, in the region where H$_2$ is more abundant than H (i.e., where H$_2$ becomes the main collision partner, indicated with vertical dashed lines in Fig.~\ref{fig:PDR-models}). Integrating from the layer where $n(\rm H_2)\ge n(\rm H)$ to the far edge of the slab, we get column densities of about $2\times 10^{15}$~cm$^{-2}$ for C$_{2}$H and $10^{15}$~cm$^{-2}$ for C$_3$H$_2$. That is, about three times and twenty five times larger than what we found from the rotational diagrams.

\begin{figure*}[htp]
  \centering
 \subfigure{\includegraphics[width=90mm]{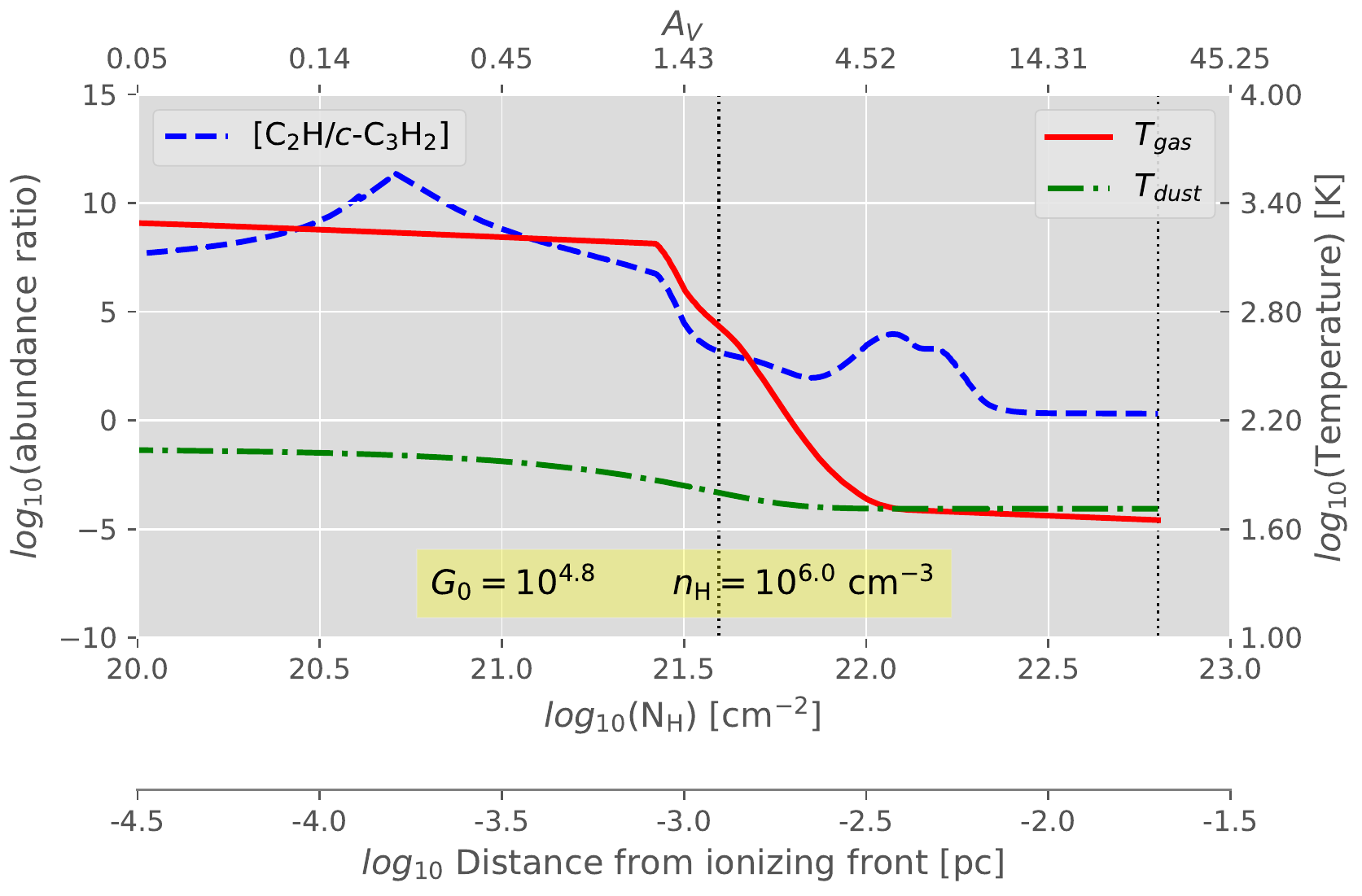}}\quad
 \subfigure{\includegraphics[width=90mm]{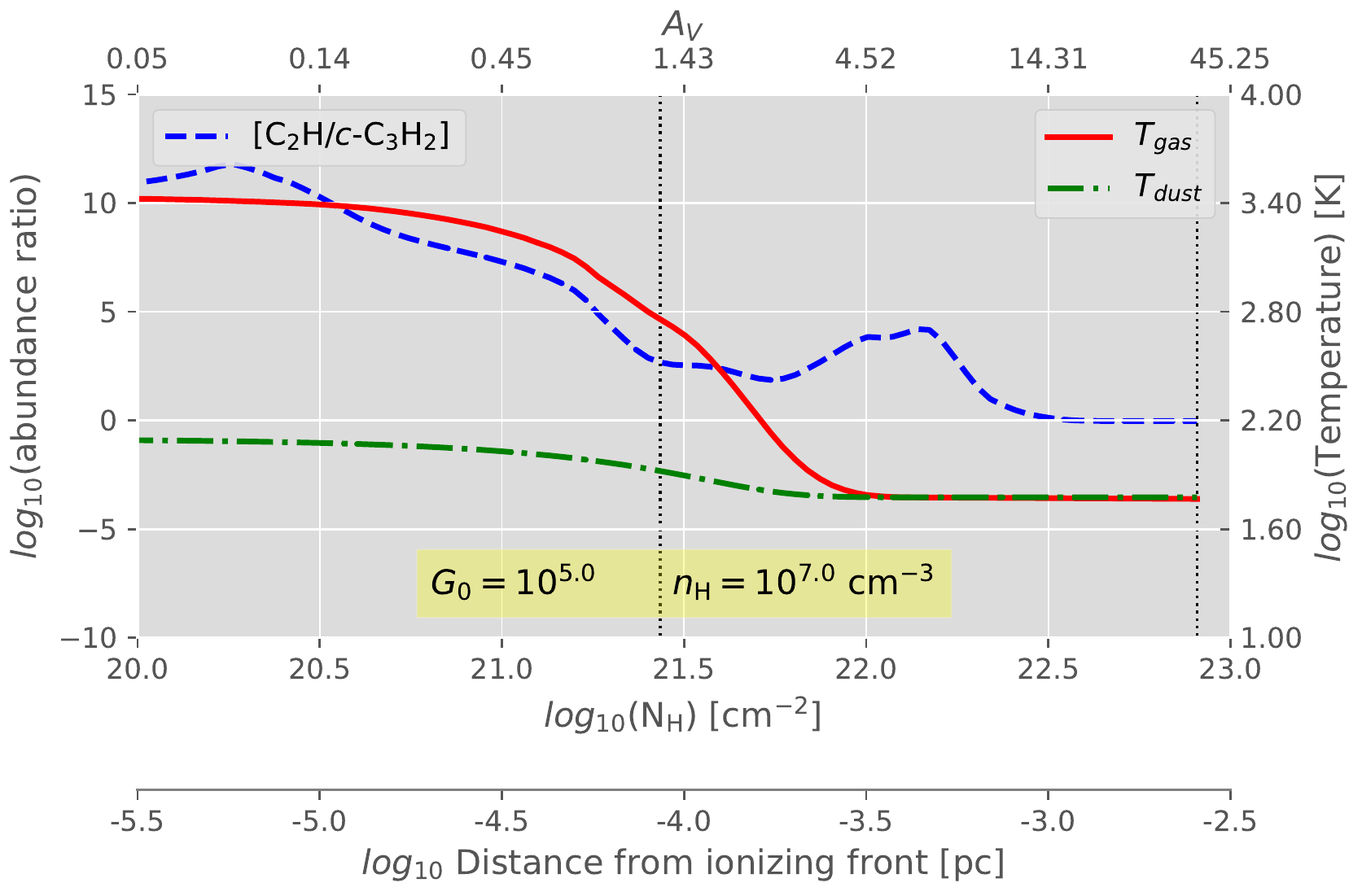}}

  \caption{The $\rm [C_{2}H/C_3H_2]$ abundance ratios from the PDR models. The left ordinate corresponds to the abundance ratio while the right ordinate corresponds to the gas and dust temperatures. Other labels are as in Fig.~\ref{fig:PDR-models}.}
  \label{fig:PDR-abundance-ratios}
\end{figure*}

For the lowest density model ($n_{\rm H} = 10^6$~cm$^{-3}$) we get relative abundances between $3\times 10^{-9}$ and $2\times 10^{-8}$ for C$_{2}$H and between $2\times 10^{-12}$ and $10^{-8}$ for C$_3$H$_2$, in the region where $n(\rm H_2)$ is dominant (i.e., for $A_{\rm V}$ between 1.8 and 28.6 mag, indicated by the vertical dashed lines in Fig.~\ref{fig:PDR-models}). The associated column densities are in the order of $10^{15}$~cm$^{-2}$ for C$_{2}$H and $4\times 10^{14}$~cm$^{-2}$ for C$_3$H$_2$. That is, between factors three and ten larger than what we found from the rotational diagrams for each species.\

The difference between the observed column densities and those predicted by the homogeneous PDR models can arise from the unresolved clumpiness in the gas, which is not taken into account in the calculations based on our observations. Hence, the column densities estimated from observations should be considered lower limits of the actual column densities of the species.\ 
The slab depth dependent $\rm [C_{2}H/C_3H_2]$ abundance ratios are shown in Fig.~\ref{fig:PDR-abundance-ratios} for the same models presented in Fig.~\ref{fig:PDR-models}. The C$_{2}$H molecule is about ten orders of magnitude more abundant over most of the slab's length for all the cases studied. However, the abundance ratio decreases to just about factor two in the lower density model ($n_{\rm H} = 10^6$~cm$^{-3}$) toward the far edge of the slab. In the higher density model ($n_{\rm H} = 10^7$~cm$^{-3}$) instead the abundance ratio turns around as the gas cools down and $\rm C_3H_2$ becomes about four times more abundant than C$_{2}$H. Gas temperatures reached at the far edge of the slab are about 45~K for the lower density model and about 60~K for the higher density model. Considering all the region of the slab where H$_2$ is the most abundant species, the average gas temperatures are about 100~K and 120~K for the lower and higher density PDR models, respectively. These average temperatures are similar to the values (100--150~K) we \citep{refId0} obtained for this region of M8.\ 

Despite not considering surface grain chemistry or mechanisms about PAH destruction due to UV radiation, the PDR models lead to temperatures comparable to our estimates based on the observations after using the same density values found with the non-LTE radiative transfer models. This is in favor of the argument that gas-phase chemistry is able to roughly explain the abundance of small hydrocarbons in high-UV flux PDRs. 

\section{Discussion}
\subsection{Physical conditions probed by two velocity components}
Two spectrally resolved velocity components are seen in the $N$ = 5 $\to$ 4 transition of C$_{2}$H (Fig.~1) and are very prominent in all transitions of c-C$_{3}$H$_{2}$ (Fig.~2). These two components probe regions of different temperatures and densities. As mentioned before, the low velocity component (2--8~km~s$^{-1}$) is a part of the gas in the foreground veil and the high velocity component (8--15~km~s$^{-1}$) is a part of the gas close to Her 36 and the background molecular cloud. Owing to the prominent foreground veil feature in c-C$_{3}$H$_{2}$, it seems to be a better probe of the warm PDR as compared to C$_{2}$H.\

The rotational diagram analysis performed in Sect.~4.2, provides us with the column densities and abundances of the low and high velocity components of C$_{2}$H and c-C$_{3}$H$_{2}$. Lower column densities and abundances are found in the foreground veil compared to the high column densities and abundances obtained in the gas close to Her 36 and in the background molecular cloud. As described in Sect.~4.3, the non-LTE RADEX modeling provides us with H$_{2}$ volume density estimates. The low velocity component (corresponding to the foreground veil) probes the high density gas in the range of 5 $\times$ 10$^{5}$--5 $\times$ 10$^{6}$~cm$^{-3}$. While, the high velocity component (corresponding to the gas close to Her 36 and in the background molecular cloud) probes both high and low density gas in the range of 5 $\times$ 10$^{4}$--5 $\times$ 10$^{6}$~cm$^{-3}$.  

\subsection{Comparison with the Orion Bar PDR}

The Orion Bar, an elongated structure in the Orion Nebula (M43), is the most-studied dense PDR (\citealt{1993Sci...262...86T}, \citealt{1997ARA&A..35..179H} \& \citealt{2000A&A...364..301W}). In particular, it is an interesting source to study small hydrocarbon chemistry in ISM. \cite{1995A&A...294..792H}, \cite{2000A&A...364..301W}; and \cite{2017A&A...598A...2A} described the geometry of the Orion Bar where the prominent O stars in the Trapezium illuminate the PDR, whose orientation changes from face-on to an edge-on relative to the line of sight and which appears to be located at the edge of the \hii\ blister \citep{2012A&A...538A..12P}. Recent observational studies of small hydrocarbons toward this PDR have been reported by \citet{2015A&A...575A..82C}, who studied the chemistry and spatial distribution of C$_{2}$H, C$_{4}$H, c-C$_{3}$H$_{2}$, c-C$_{3}$H and by \citet{2015A&A...578A.124N}, who constrained the physical conditions of the C$_{2}$H (high transitions from $N$ = 6 $\to$ 5 to 10 $\to$ 9) emitting gas toward the Orion Bar.\

In Table~\ref{m8_orion}, we compared the C$_{2}$H and c-C$_{3}$H$_{2}$ column densities toward M8 that we observe and obtain from modeling with those reported toward the Orion Bar by \citet{2015A&A...575A..82C}. We find the observed column densities to be similar in both PDRs, which was expected as we also found such similarities in the H$_{2}$ volume density, in the kinetic temperature and in the CO and \cii\ luminosities \citep{refId0}.    

On the other hand, we found significant differences between the observed and the modeled column density of the C$_3$H$_2$ molecule. In the case of Orion it is about one order of magnitude \textit{lower} than estimated from the observations. In the case of M8 the modeled column density is between one and two orders of magnitude \textit{larger} than estimated from observations, which is in line with the assumption that column densities we have determined should be considered lower limits. This suggests that there might be a significant difference between the results obtained with the Meudon PDR model used by \citet{2015A&A...575A..82C} and our PDR model based on \citet{2005A&A...436..397M}. As shown in Appendix~\ref{fig:PDR-orion-models}, we ran our models for the Orion Bar with the same input parameters ($n_{\rm H} = 4 \times 10^{6}$~cm$^{-3}$ and $G_0 = 1.17 \times 10^{4}$) as used by \citet[Fig.~17, right panel]{2015A&A...575A..82C}. We believe that the different chemical networks used in each model that lead to the formation and destruction of the C$_3$H$_2$ molecule, are mainly responsible for their different column density predictions. 
 

\begin{table}
\centering
\begin{threeparttable}
\caption{Column densities of C$_{2}$H and c-C$_{3}$H$_{2}$ in M8 and the Orion Bar}
\begin{tabular}{c c c c c}
\hline\hline
 \noalign{\smallskip}
Species & \multicolumn{4}{c}{Column densities log$_{\rm 10}$($N$) (cm$^{-2}$)}\\
        & \multicolumn{2}{c}{Observed\tnote{a}} & \multicolumn{2}{c}{PDR model\tnote{b}}\\
        & M8 & Orion & M8 & Orion\\
\hline
\noalign{\smallskip}
C$_{2}$H     & 14.6 & 14.6 &15   & 14.1--14.8\\
c-C$_{3}$H$_{2}$ & 13.5 & 13.1 & 14.6--14.9&12.2--12.9 \\

\hline
 \noalign{\smallskip}
\end{tabular}
\begin{tablenotes}
            \item[a]\small Values for M8, calculated from Table~\ref{rot_dia} and for Orion Bar, obtained from \citet[Table~7]{2015A&A...575A..82C}.
            \item[b]\small Values for M8, obtained from the modeling results presented in Sect.~4.4 and for Orion Bar, obtained from \citet[Table~7]{2015A&A...575A..82C}.
\end{tablenotes}

\label{m8_orion}
\end{threeparttable}
\end{table}





\section{Conclusions}
In this paper, we report APEX and IRAM~30~m observations of the $N$ = 1 $\to$ 0, 3 $\to$ 2 and 5 $\to$ 4 transitions of C$_{2}$H and of the $J$ = 2 $\to$ 1, 6 $\to$ 5, 7 $\to$ 6 and 8 $\to$ 7 transitions of c-C$_{3}$H$_{2}$ toward M8. We presented the spectra of 18 observed hfs components of C$_{2}$H and 5 observed transitions of ortho and para species of c-C$_{3}$H$_{2}$ along with the velocity integrated intensity maps of the brightest $N$ = 1 $\to$ 0, $J$ = 3/2 $\to$ 1/2, $F$ = 2 $\to$ 1 and $N$ = 3 $\to$ 2, $J$ = 7/2 $\to$ 5/2, $F$ = 4 $\to$ 3 transitions of C$_{2}$H.\

As discussed in Sect.~4.2, we inferred column densities ranging from 10$^{12}$ to 10$^{14}$~cm$^{-2}$. C$_{2}$H is more abundant than c-C$_{3}$H$_{2}$ with an abundance $\sim$ 10$^{-8}$. The C$_{3}$H$_{2}$ ortho/para ratio is calculated to be $\sim$ 2.2 for low velocity component (corresponding to the gas in the warm foreground veil) and 3.3 for high velocity component (corresponding to the gas near Her 36 and the background molecular cloud). We also expect c-C$_{3}$H$_{2}$ to be a better probe than C$_{2}$H for the warm PDR owing to its well defined low velocity component feature (seen in Fig.~2).\

Using non-LTE RADEX modeling (Sect.~4.3), we constrained the H$_{2}$ volume densities of the hydrocarbon emitting gas to 5 $\times$ 10$^{4}$ to 5 $\times$ 10$^{6}$~cm$^{-3}$. In Sect.~4.4, we compared the observed column densities with updated PDR models and they match the observed C$_{2}$H column densities reasonably well (by a factor $\sim$~3) but predict higher values for c-C$_{3}$H$_{2}$ (by a factor $\sim$ 10--25). This discrepancy might arise from clumpy gas structure that we do not resolve with our observations. Therefore, the column densities derived from our data should be considered to be lower limits. The spatial distribution of PAH emission does not follow the C$_{2}$H emission. This is consistent with PDR models, which do not require PAH fragmentation to explain the observed column densities of C$_{2}$H and c-C$_{3}$H$_{2}$ in M8. 
       
\bibliography{main}
\bibliographystyle{aa}
\clearpage
\onecolumn
\begin{appendix}

\section{Identified lines of hydrocarbons}
 
\begin{table*}[h]
\centering
\begin{threeparttable}
\caption{Line intensities of the observed hfs transitions of C$_2$H line taken from CDMS.}

\begin{tabular}{c c c c c c }
 \hline\hline
 \noalign{\smallskip}
       Transition & Frequency & S$_{\rm ij}\tnote{a}$ & T$_{\rm mb}$\tnote{b}& \multicolumn{2}{c}{Relative Intensities} \\
 \hline
 \noalign{\smallskip}
  (J,F)$_{N=p+1}$ $\rightarrow$ (J,F)$_{N=p}$ & GHz & & K & LTE\tnote{c} & Obs.\tnote{d} \\
  \hline
   \noalign{\smallskip}
    \multicolumn{6}{c}{C$_{2}$H $N = 1 - 0$ with IRAM 30m/EMIR}\\
 \hline
 \noalign{\smallskip}
 (3/2,1) $\rightarrow$ (1/2,1) & 87.284 & 0.17 & 0.16 & 0.042 & 0.055 \\
 (3/2,2) $\rightarrow$ (1/2,1) & 87.316 &1.67  & 1.10 &0.416 & 0.392 \\ 
 (3/2,1) $\rightarrow$ (1/2,0)  & 87.328 & 0.83 & 0.57 &0.207 & 0.204 \\
 (1/2,1) $\rightarrow$ (1/2,1)  & 87.401 & 0.83 & 0.58 &0.208& 0.208 \\
 (1/2,0) $\rightarrow$ (1/2,1) & 87.407 & 0.33 & 0.25 &0.083& 0.088\\
 (1/2,1) $\rightarrow$ (1/2,0)  & 87.446 & 0.17 & 0.15 &0.042 & 0.052\\ 
 \hline
 \noalign{\smallskip}
 \multicolumn{6}{c}{C$_{2}$H $N = 3 - 2$ with APEX/PI230}\\
 \hline
 \noalign{\smallskip}
 (7/2,3) $\rightarrow$ (5/2,3) & 261.978 & 0.110 & 0.006  &0.009 & 0.013\\
 (7/2,4) $\rightarrow$ (5/2,3) & 262.004 & 3.857 & 1.45 & 0.321 & 0.290\\
 (7/2,3) $\rightarrow$ (5/2,2) & 262.006 &2.886 & 1.37 & 0.240 &0.270\\
 (5/2,3) $\rightarrow$ (3/2,2) & 262.064 &2.755 & 1.00 &0.230 &0.197 \\
 (5/2,2) $\rightarrow$ (3/2,1) & 262.067 &1.800 & 0.948 & 0.150& 0.186\\
 (5/2,2) $\rightarrow$ (3/2,2) & 262.078 &0.242 & 0.102 &0.020 &0.020 \\
 (5/2,3) $\rightarrow$ (5/2,3) & 262.208 &0.223 & 0.123 &0.019 &0.024 \\
 (5/2,2) $\rightarrow$ (5/2,2) & 262.250 &0.091 & 0.005& 0.008& 0.010 \\
 \hline
 \noalign{\smallskip}
 \multicolumn{6}{c}{C$_{2}$H $N = 5 - 4$ with APEX/FLASH$^+$}\\
 \hline
 \noalign{\smallskip}
 (11/2,6) $\rightarrow$ (9/2,5)& 436.661 &5.910 & 0.8 & 0.298 & 0.296 \\
 (11/2,5) $\rightarrow$ (9/2,4)& 436.662 &4.921 & 0.7 & 0.248 & 0.259 \\
 (9/2,5) $\rightarrow$ (7/2,4)& 436.723 &4.837 & 0.665 & 0.245 & 0.246 \\
 (9/2,4) $\rightarrow$ (7/2,3)& 436.724 &3.890 & 0.535 & 0.196 & 0.198 \\
  \hline
 \noalign{\smallskip}
 
  \end{tabular}
  \begin{tablenotes}
            \item[a]\small Theoretical line strengths.
            \item[b]\small Observed main beam temperatures.
            \item[c]\small S$_{\rm ij}$/$\sum$S$_{\rm ij}$, assuming optically thin transitions ($\tau$ $<$ 1)
            \item[d]\small T$_{\rm mb_{ij}}$/$\sum$T$_{\rm mb_{ij}}$
\end{tablenotes}

  \label{rel_int_cch}
  \end{threeparttable}
\end{table*}
\twocolumn

\section{Spatial distribution of \ci\, and CO $J$ = 6 $\rightarrow$ 5 relative to C$_2$H $N$ = 1 $\rightarrow$ 0}

As shown in Fig.~B.1, \ci\ emission (in color) peaks close to Her 36 and has a bright extended emission toward the north-west of it. This extended emission is also traced by CO $J$ = 6 $\to$ 5 emission (white contours) as well as by C$_2$H $N$ = 1 $\rightarrow$ 0 emission (black contours).

\begin{figure}[htp]
  \centering
 \subfigure{\includegraphics[width=80mm]{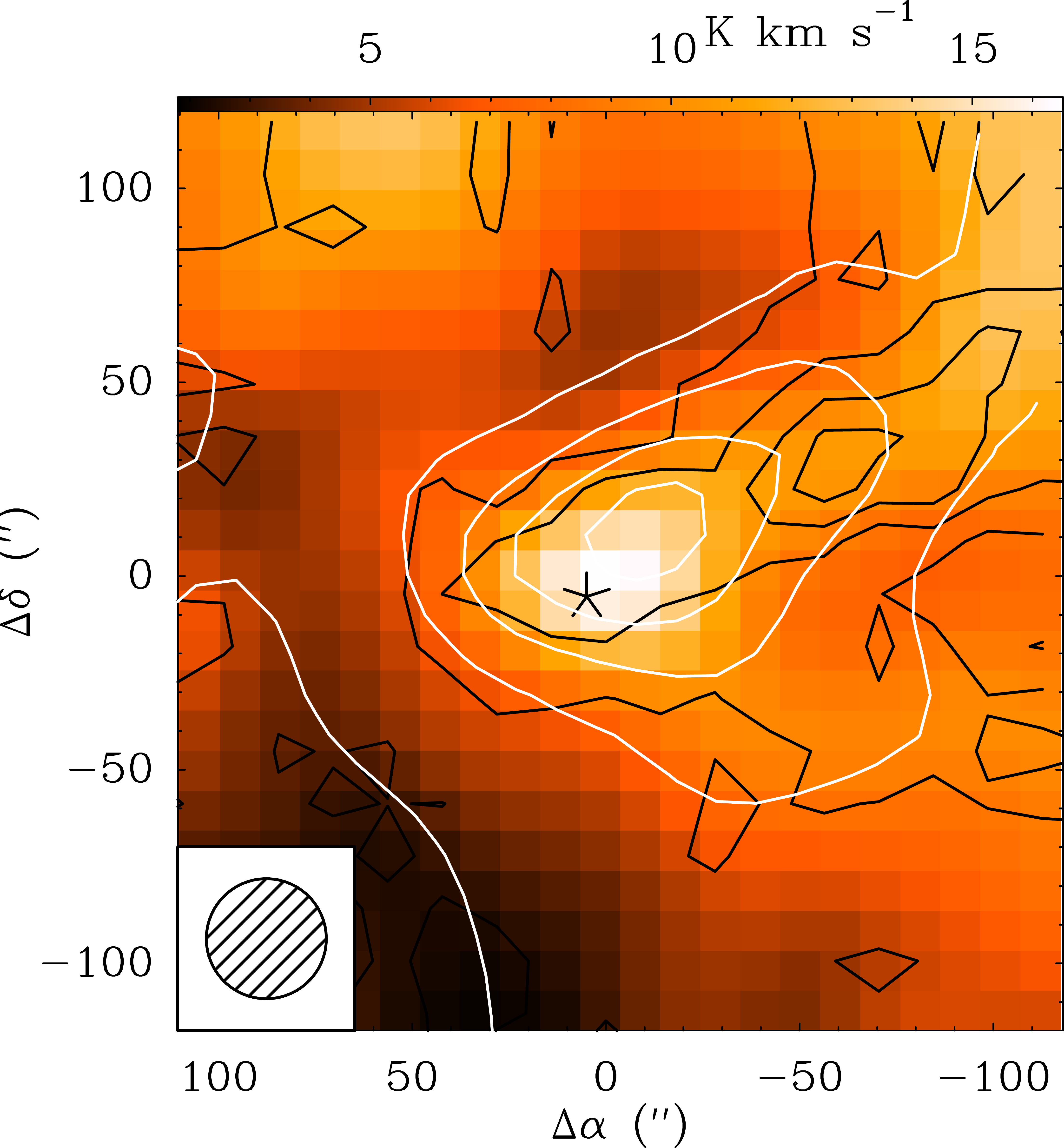}}

  \caption{Colour map of velocity integrated intensity of the \ci\, 609~$\mum$ line overlaid with velocity integrated intensity map contours of C$_2$H (in black) and CO $J$ = 6 $\rightarrow$ 5 (in white). Her 36 is the central position ($\Delta\alpha$ = 0, $\Delta\delta$ = 0) at R.A.(J2000) = 18\textsuperscript{h}03\textsuperscript{m}40.3\textsuperscript{s} and Dec.(J2000) = --24\degree22$\arcmin$43$\arcsec$, marked with an asterisk. For both C$_2$H and CO $J =$ 6 $\rightarrow$ 5, the contour levels are 10\% to 100\% in steps of 20\% of the peak emission. All maps were convolved to the same resolution of 30$''$.}
  \label{fig:ci-co-c2h}
\end{figure}

\section{PDR modeling results for Orion Bar conditions}
\begin{figure}[htp]
  \centering
 \subfigure{\includegraphics[width=80mm]{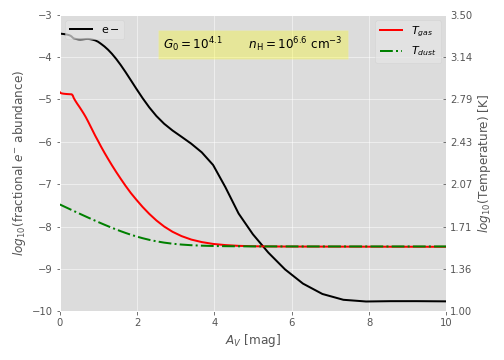}}
 \subfigure{\includegraphics[width=80mm]{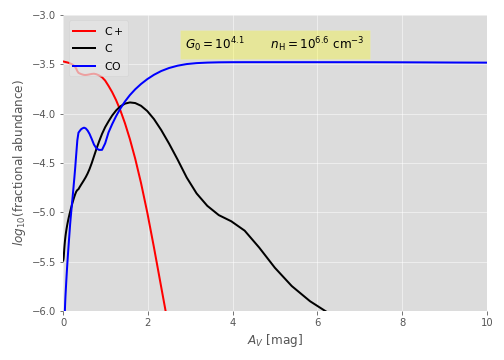}}
 \subfigure{\includegraphics[width=80mm]{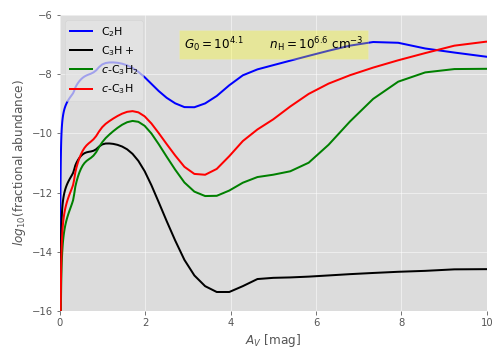}}
  \caption{Results of our PDR modeling for a high density clump in the Orion Bar ($n_{\rm H}$ =  4 $\times$ 10$^6$  cm$^{-3}$  and $G_{\rm 0}$ =  1.17 $\times$  10$^4$). In all panels, the abscissa gives visual extinction. The top panel shows the dependence of the electron abundance (black solid line/ left ordinate), gas temperature (red solid line/ right ordinate) and dust temperature (red line, green dash-dot/ right ordinate). The middle panel plots the fractional abundances of ionized and neutral carbon and CO as red, black and blue lines, respectively. The bottom panel plots the fractional abundances of C$_2$H, C$_2$H$^+$, c-C$_3$H$_2$ and c-C$_3$H as ionized and neutral carbon and CO as blue, black, green and red lines, respectively.}
  \label{fig:PDR-orion-models}
\end{figure}

To investigate the different c-C$_{3}$H$_{2}$ column density values obtained by our PDR models \citep{2005A&A...436..397M} for M8 and by the Meudon code for the Orion Bar, we ran our models using the same input parameters as \citet[Fig.~17, right panel]{2015A&A...575A..82C} for their high density clump ($n_{\rm H} = 4 \times 10^{6}$~cm$^{-3}$ and $G_0 = 1.17 \times 10^{4}$). The modeling results are presented in Fig.~\ref{fig:PDR-orion-models}. Clearly our models do not produce the same results as the Meudon code \citep[Fig.~17, right panel]{2015A&A...575A..82C}. The differences in both models lie in the initial abundances of various species, in their evolution profiles and in the temperature profiles throughout the slab (A$_v$ = 0 to 10). Differences may arise due to distinct chemical networks used in both models. The Meudon code run by \citet{2015A&A...575A..82C} uses a total of 130 species and 2800 gas-phase reactions. The \citet{2005A&A...436..397M} model, on the other hand, uses a total of 309 species and 4453 gas-phase reactions. The different initial abundances of the neutral and ionized carbon at the edge (A$_v$ = 0) of our PDR model and that of the Meudon code may also contribute to the higher column density prediction of c-C$_3$H$_2$. Furthermore, it is worth mentioning that the Meudon code includes the updated state-to-state reactions of vibrationally excited H$_{2}$ with \cii\,, \oi\,, OH \citep{2010ApJ...724L.133A}. It also has upgraded the carbon-bearing species network and used the most recent branching ratios for ion-molecule, neutral-neutral, dissociative recombination, and charge exchange reactions for carbon chains and hydrocarbon species as described in \citet{2013ApJ...771...90C}. These updates are missing in our PDR models. The actual impact of these exchange reactions cannot be measured until the PDR model by \citet{2005A&A...436..397M} is updated.    

\end{appendix}

\end{document}